\documentclass[lettersize,journal]{IEEEtran}
\usepackage{amsmath,amsfonts}
\usepackage{algorithmic}
\usepackage{algorithm}
\usepackage{array}
\usepackage[caption=false,font=normalsize,labelfont=sf,textfont=sf]{subfig}
\usepackage{textcomp}
\usepackage{stfloats}
\usepackage{url}
\usepackage{verbatim}
\usepackage{graphicx}
\usepackage{cite}
\usepackage{booktabs} 
\usepackage{siunitx}  
\usepackage{balance}
\begin{document}
	\title{HADUA: Hierarchical Attention and Dynamic Uniform Alignment for Robust Cross-Subject Emotion Recognition}
	\author{Jiahao Tang, \textit{Student Member, IEEE},  Youjun Li, Yangxuan Zheng, Xiangting Fan, Siyuan Lu, Nuo Zhang and Zi-Gang Huang$^{*}$ 
		\thanks{This work was supported by the Natural Science Foundation of China (No.11975178), Natural Science Basic Research Program of Shaanxi (No. 2023-JC-YB-07) and Shaanxi Fundamental Science Research Project for Mathematics and Physics (Grant No.22JSQ037)}
		\thanks{Jiahao Tang, Youjun Li, Yangxuan Zheng, Xiangting Fan, Siyuan Lu, Nuo Zhang and Zi-Gang Huang are affiliated with the Key Laboratory of Biomedical Information Engineering, Ministry of Education, Institute of Health and Rehabilitation Science, School of Life Science and Technology, Xi'an Jiaotong University. They are also associated with the Key Laboratory of Neuro-Informatics \& Rehabilitation Engineering, Ministry of Civil Affairs, Xi'an, Shaanxi, China, as well as the Research Center for Brain-Inspired Intelligence, Xi'an Jiaotong University.}}

\maketitle

\begin{abstract}
	Robust cross-subject emotion recognition from multimodal physiological signals remains a challenging problem, primarily due to modality heterogeneity and inter-subject distribution shift. To tackle these challenges, we propose a novel adaptive learning framework named Hierarchical Attention and Dynamic Uniform Alignment (HADUA). Our approach unifies the learning of multimodal representations with domain adaptation. First, we design a hierarchical attention module that explicitly models intra-modal temporal dynamics and inter-modal semantic interactions (e.g., between electroencephalogram(EEG) and eye movement(EM)), yielding discriminative and semantically coherent fused features. Second, to overcome the noise inherent in pseudo-labels during adaptation, we introduce a confidence-aware Gaussian weighting scheme that smooths the supervision from target-domain samples by down-weighting uncertain instances. Third, a uniform alignment loss is employed to regularize the distribution of pseudo-labels across classes, thereby mitigating imbalance and stabilizing conditional distribution matching. Extensive experiments on multiple cross-subject emotion recognition benchmarks show that HADUA consistently surpasses existing state-of-the-art methods in both accuracy and robustness, validating its effectiveness in handling modality gaps, noisy pseudo-labels, and class imbalance. Taken together, these contributions offer a practical and generalizable solution for building robust cross-subject affective computing systems.
\end{abstract}


\section{Introduction}
\label{sec:introduction}
Emotion recognition constitutes a fundamental research challenge in the fields of affective computing and human-computer interaction, as emotional states play a critical role in shaping human cognition, behavior, and decision-making \cite{911197,YOO2006345}. Accurate emotion recognition not only contributes to improve individuals well-being but also holds significant importance for enhancing system performance in task-oriented applications such as adaptive human-machine systems and intelligent interaction interfaces. In recent years, multi-modal physiological signals have gained widespread attention due to their capacity to provide objective and interpretable indicators of emotional responses \cite{RAHMAN2021104696}. Among them, electroencephalogram (EEG) signals reflect neural activity characteristics from the central nervous system, while eye movement (EM) signals capture overt behavioral cues, including fixation patterns and pupil dynamics. Existing studies have shown that EEG and EM signals exhibit significant complementarity and synergy in emotion recognition tasks \cite{9395500,6944757,8283814}.

Despite the rich information offered by multi-modal physiological signals for emotion recognition, current research faces core challenges in practical applications: how to achieve stable generalization of multi-modal emotion recognition models under cross-subject conditions \cite{10.3389/fnins.2018.00162,10.1145/3664647.3681579}. This challenge is primarily manifested in two aspects. On one hand, multi-modal heterogeneity inherently exists. EEG and EM signals differ significantly in temporal scales, statistical distributions, and physiological origins, making direct alignment or simple fusion between different modalities difficult \cite{10575932}. On the other hand, cross-subject distribution inconsistency is particularly prominent. Due to individual differences in neural activity patterns and behavioral responses, data from different subjects often exhibit significant distribution shifts, severely violating the independent and identically distributed (i.i.d.) assumptions underlying conventional models \cite{pan2009survey,kouw2019review}. These two issues are intertwined, making cross-subject multi-modal emotion recognition an extremely demanding task.

To address these challenges, existing studies have mainly explored two relatively independent directions. To tackle multi-modal heterogeneity, numerous works have focused on designing multi-modal fusion strategies, such as feature-level concatenation, correlation alignment, tensor fusion, and attention-based fusion-which achieve certain success under single-subject or controlled conditions by modeling and integrating different modalities at the representation level \cite{10.5555/2832249.2832411,8283814,9395500}. For the problem of cross-subject distribution difference, transfer learning and domain adaptation methods have been widely adopted in the emotion recognition field, alleviating performance degradation caused by individual differences by aligning feature distributions, marginal distributions, or conditional distributions between source and target domains \cite{pan2009survey,wu2020transfer,8882370}. Among them, methods based on joint alignment of marginal and conditional probability distributions have shown strong adaptability in EEG-based emotion recognition \cite{10506974,10819285}.

However, when multi-modal fusion and cross-subject domain adaptation are considered simultaneously, existing methods still face several key limitations. First, under cross-subject adaptation frameworks, multi-modal heterogeneity is no longer merely a challenge at the representation fusion level but directly impairs the reliability of pseudo-labels in the target domain. When multi-modal representations fail to explicitly model intra-modal structures and inter-modal semantic relationships, inconsistencies in discriminative information across modalities are easily amplified during the prediction stage, thereby introducing systematic bias into target domain pseudo-label \cite{HANGLOO2025130827, electronics12153325}. Second, pseudo-label learning in the target domain generally faces a quantity-quality trade-off: incorporating numerous pseudo-labels in early training stages tends to introduce noise, whereas over-reliance on high-confidence filtering results in insufficient supervision, making it difficult to cover all emotion categories \cite{sohn2020fixmatchsimplifyingsemisupervisedlearning,chen2023softmatchaddressingquantityqualitytradeoff}. Third, due to the inherent biases in model predictions themselves, pseudo-labels in the target domain often exhibit severe class imbalance, which not only affects classification performance but also undermines the reliable estimation of class-conditional statistics, thereby weakening the effectiveness of conditional probability distribution alignment methods \cite{10506974,10509712}. The superposition of these issues severely constrains the stability and generalization ability of cross-subject multi-modal emotion recognition models.

To overcome the aforementioned challenges, this paper proposes a novel cross-subject multi-modal emotion recognition framework, termed HADUA, which enables robust emotion recognition from EEG and EM signals. The framework jointly addresses multi-modal heterogeneity and pseudo-label unreliability, significantly enhancing the model's generalization ability under domain shifts. The main contributions of this work are summarized as follows:

1) Hierarchical Attention-based Multi-modal Fusion Method  
We propose a hierarchical attention fusion framework to explicitly model intra-modal and inter-modal dependencies in EEG and EM signals. Modality-specific self-attention modules extract robust temporal representations, and a cross-modal attention mechanism achieves semantically guided interactions, effectively alleviating noise introduced by modality heterogeneity and avoiding spurious correlations.

2) Confidence-aware Gaussian-weighted Domain Adaptation Strategy  
To address the quantity-quality trade-off in pseudo-label learning, we design a truncated Gaussian weighting mechanism that dynamically assigns pseudo-label weights based on target domain prediction confidence, suppressing the influence of low-quality pseudo-labels while retaining effective supervision signals, thereby promoting stable alignment of marginal and conditional distributions.

3) Uniform Alignment Mechanism for Class-level Pseudo-label Balancing  
To mitigate class imbalance in target domain pseudo-labels, we introduce a Uniform Alignment mechanism that applies soft constraints to regulate class-level pseudo-label distributions, ensuring balanced contributions from each emotion category during conditional distribution alignment and further improving the model's generalization performance.

Experimental results on multiple cross-subject emotion recognition benchmarks demonstrate that the proposed framework outperforms state-of-the-art methods significantly, confirming its effectiveness and robustness. 
The remainder of this paper is organized as follows: Section II reviews the related work. Section III details the design of the whole framework. Then, the performances of the proposed method are illustrated in experiments and compared with other conventional methods in section IV. Finally, in Section V, some of the distinctive features of this investigation are highlighted.

\begin{table}[h!]
	\begin{center}
		\caption{Notations and descriptions used in this paper.}\label{tab1}
		\setlength{\tabcolsep}{1mm}
		\begin{tabular}[t]{l|c}
			\hline
			Notation & Description \\
			\hline
			$D_s=\left\{ x_{s}^{l},y_{s}^{l} \right\} _{i=1}^{N_{}}$ &Source domain\\
			$D_t=\left\{ x_{t}^{u} \right\} _{i=1}^{Nu}$&Target domain \\
			$f()$& Feature extractor \\
			$\hat{y}_{s}^{l}$ & Source domain predict labels \\
			$\hat{y}_{t}^{u}$ & Target generate pseudo labels\\
			$K$ & Gaussian Kernel function \\
			$RKHS$ & Reproducing Kernel Hilbert Space\\
			$SGD$  & Stochastic Gradient Descent\\
			$ReLU$ & Rectified Linear unit activation function\\
			\hline
		\end{tabular}
	\end{center}
\end{table}

\section{Related Work}
\subsection{Emotion Recognition With EEG and EM Signals}
Multimodal fusion methods, which leverage the complementary information inherent in multimodal data, have been widely adopted in emotion recognition to enhance performance \cite{10.3389/fncom.2016.00085,8283814,10731546}. Prior research has explored various methods combining EEG and EM features for emotion recognition. For instance, Lu et al. investigated multiple fusion strategies—including feature-level concatenation, MAX fusion, SUM fusion, and fuzzy integral fusion—to integrate EEG and EM signals \cite{10.5555/2832249.2832411}. In deep learning-based approaches, the quality of modality representations is crucial for model performance. These methods can be broadly categorized into two paradigms: joint representation learning and coordinated representation learning \cite{8882370}. The former maps EEG and EM signals into a unified feature space designed to represent emotional information consistently, exemplified by models like BDAE \cite{Liu2016} and MFFNN \cite{10.3389/fnins.2023.1234162}. The latter projects the signals into a constrained coordinated hyperspace by maximizing inter-modal correlations, as seen in CCA-based methods \cite{9395500,Qiu2018}. However, simple fusion strategies often fail to capture high-order relationships between modalities, resulting in incomplete representations. Moreover, enforcing rigid alignments risks losing modality-specific details, which can degrade performance in cross-subject scenarios.

Recent advances in multimodal fusion have employed more sophisticated techniques to integrate EEG and EM signals. Zhou et al. proposed a subjective and objective feature fusion method using element-wise multiplication to combine EEG and EM features \cite{ZHOU2022108889}. Lian et al. employed similarity-based cross-modal alignment to enhance modality interactions \cite{10643252}. Li et al. introduced UMAP, a mixture-of-experts framework that flexibly combines modality-specific and shared experts. It employs multi-task pretraining—including contrastive learning, matching, and generation—with tailored attention masks to fuse EEG and EM, maintaining robustness even with missing modalities \cite{10.1145/3746027.3755459}. While these methods effectively model cross-modal interactions and project features into a latent space, they often overlook hierarchical and dynamic relationships, as well as nonlinear dependencies and temporal disparities, leading to loss of modality-specific details and limited noise robustness, highlighting the need for more comprehensive fusion strategies such as our proposed hierarchical attention-based approach.

\subsection{Domain Adaptation for Cross-Subject Emotion Recognition}
Affective brain-computer interfaces that utilize EEG and EM signals have attracted considerable research interest due to their potential in healthcare and emotion-aware systems. However, the high inter-subject and cross-session variability in EEG patterns, recording setups, and non-stationary signal distributions—poses a major obstacle to developing robust emotion classification models. Domain adaptation methods have consequently emerged as a key solution to bridge the gap between source and target domains, generally falling into two primary categories.

The first category employs advanced techniques—including adversarial learning, graph-based modeling, and attention mechanisms—to align domain distributions while improving the modeling of temporal EEG dynamics and EM feature representations. For example, Li et al. proposed LGDAAN-Nets, a domain adaptation framework incorporating ConvLSTM, which integrates local and global domain discriminators to achieve cross-subject accuracies of 89.09\% on the SEED dataset and 74.12\% on SEED-IV \cite{AN2025113613}. Zhu et al. proposed CSMM, a cross-subject multi-modal emotion recognition framework that combines dynamic adversarial domain adaptation with self- and cross-attention mechanisms \cite{10938180}. By incorporating contrastive losses, the model reduces inter-modal heterogeneity and enhances semantic consistency, reaching 94.96\% accuracy on SEED and 89.82\% on SEED-IV. Despite their performance, these methods often rely on computationally intensive architectures, which increases operational costs.

The second category focuses on distribution alignment, assuming better-aligned representations improve generalization. Techniques such as MMD loss, CMMD loss, and W-distance are used to learn latent representations shared between domains. Training leverages labeled source and unlabeled target data, yet reliable pseudo-labels for the target domain are often lacking, impairing both precision and convergence speed. Recent work explores proxy-label methods to strategically combine labeled and unlabeled EEG data \cite{qiu2024review}. Zhang et al. applied post-augmentation to generate pseudo-labels, enriching the dataset \cite{9904937}. Yang et al. averaged predictions from all source classifiers to estimate target probabilities \cite{10509712}. Gong et al. introduced weighted representation alignment using a fixed-parameter SVM trained on source data to produce target pseudo-labels \cite{10506974}. However, these methods often struggle to balance pseudo-label quantity and quality \cite{chen2023softmatchaddressingquantityqualitytradeoff}. Confidence thresholding either discards low-confidence samples or introduces label noise, hampering generalization for complex emotions. Furthermore, pseudo-labeling can yield imbalanced distributions, as some emotion classes are predicted more confidently, biasing the model. Most approaches also fail to address multimodal heterogeneity and cross-subject variability jointly, reducing robustness in practical scenarios.

\section{Method}
This paper proposes a hierarchical attention and dynamic uniform alignment framework for robust cross-subject emotion recognition, as illustrated in Fig.~\ref{fig:method}. The proposed framework aims at cross-subject multimodal emotion recognition and consists of three collaborative components: multimodal feature learning, distribution alignment, and pseudo-label optimization. EEG and eye-movement signals collected from different subjects are first encoded by modality-specific feature extraction networks. A hierarchical attention mechanism is then employed to explicitly model intra-modality structural dependencies and inter-modality semantic interactions, yielding more discriminative fused representations. Based on the fused features, a shared classifier produces prediction probabilities for target-domain samples, which are treated as soft pseudo-label representations of the target domain.

Building upon this, the framework introduces a confidence-driven pseudo-label weighting mechanism to adjust target-domain pseudo-labels at the sample-level, thereby mitigating the adverse impact of low-reliability pseudo-labels during model optimization. The weighted soft pseudo-labels are utilized for conditional probability distribution alignment via CMMD, while the fused features themselves simultaneously participate in marginal distribution alignment using MMD. Through the collaborative interaction among multimodal feature learning, pseudo-label optimization, and distribution alignment, the proposed framework establishes a closed-loop cross-subject adaptive learning process.

\subsection{Multimodal Feature Extraction}
To capture complementary cognitive cues from heterogeneous biosignals, we adopt a dual-branch architecture for modality-specific feature extraction, followed by an attention-based fusion mechanism. Specifically, our framework processes EEG and eye-movement signals independently, applies self-attention within each modality, and optionally integrates cross-modality attention for deeper interaction.

\subsubsection{EEG Feature Representation}
In the domain of EEG signal analysis, a rigorous feature extraction process was implemented to distill meaningful information from the raw EEG data. The procedure began with the careful selection of specific time segments from the EEG signal, to capture temporally relevant information. Following this, the short-time Fourier transform (STFT) was employed to decompose the EEG signal into its constituent frequency bands: delta ($\delta$), theta ($\theta$), alpha ($\alpha$), beta ($\beta$), and gamma ($\gamma$) bands. This decomposition allowed for the isolation and examination of different frequency components, each of which is associated with distinct neurological and cognitive processes.

\begin{figure*}[t]
	\centering
	\includegraphics[width=\linewidth]{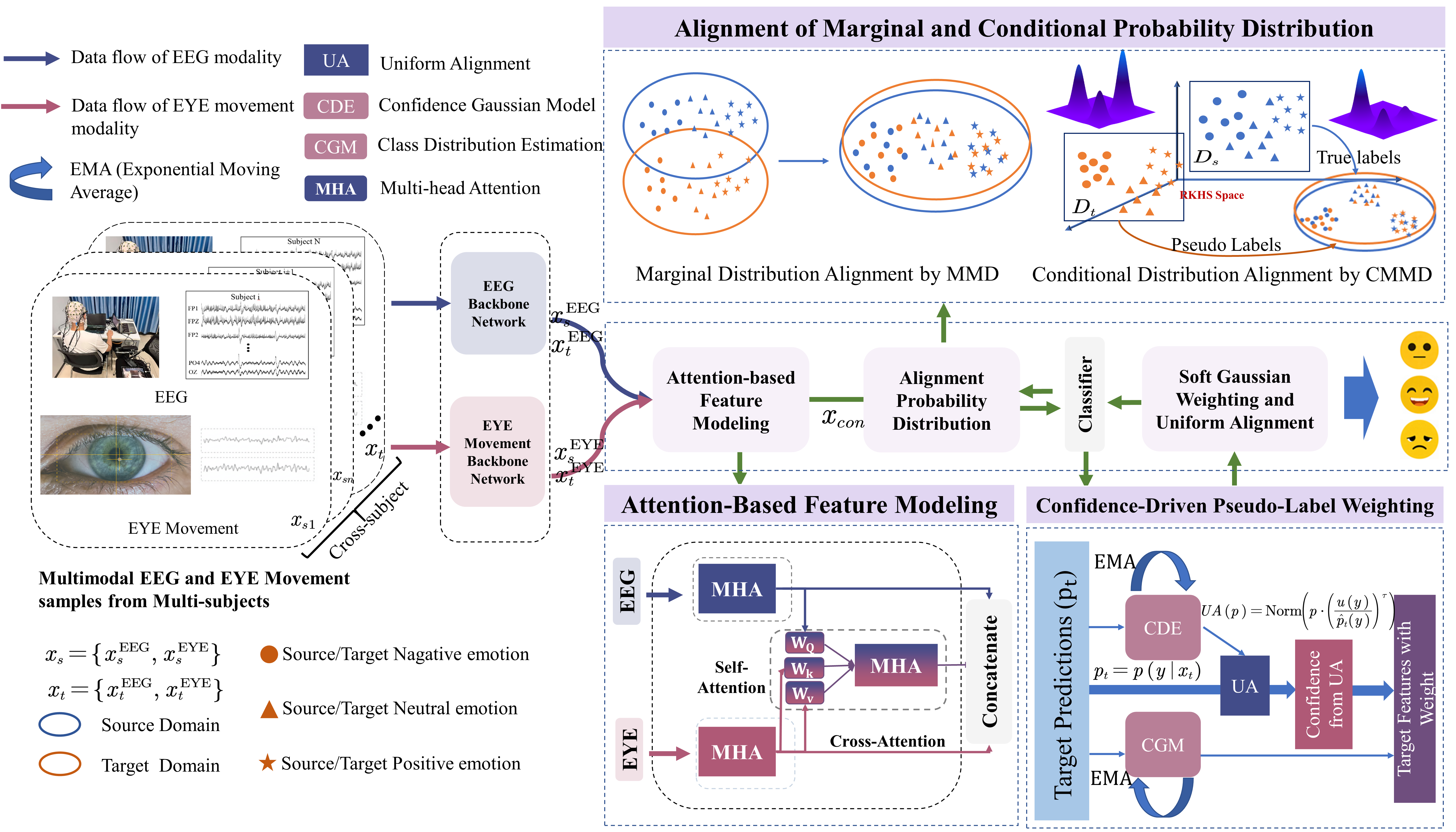}
	\caption{Overview of the proposed cross-subject multimodal emotion recognition framework. The framework mainly consists of three mutually collaborative components: (1) Attention-based multimodal feature fusion module, which employs modality-specific self-attention and cross-modal attention mechanisms to jointly model the temporal structures of EEG and eye-movement signals as well as their cross-modal semantic relationships, yielding more discriminative fused representations; (2) Multi-level distribution alignment module, which aligns the marginal distributions between the source and target domains using MMD, and further performs conditional distribution alignment via CMMD based on target-domain pseudo-labels to alleviate cross-subject domain shift; (3) Confidence-driven pseudo-label optimization module, which applies confidence-based Gaussian modeling (CDE), class distribution estimation (CGM), and a uniform distribution alignment strategy to perform sample-level weighting of target-domain pseudo-labels, thereby suppressing the adverse effects of low-reliability pseudo-labels during model optimization.}
	\label{fig:method}
\end{figure*}

Within each frequency band, the probability density function (PDF) of the signal amplitude was calculated. The PDF provides a statistical representation of the distribution of signal amplitudes within each frequency band, offering insights into the underlying characteristics of the EEG signal.

The central aspect of this feature extraction process is the computation of differential entropy, a measure that quantifies the uncertainty and complexity of the EEG signal. Differential entropy is a continuous analog of the discrete Shannon entropy and is defined as:
\begin{equation}\label{entropy}
	H\left( x \right) =\mathrm{ }-\int{p(x)\log p(x) dx}
\end{equation}

In Eq.(\ref{entropy}), $H(X)$ represents the differential entropy of the random variable $x$, which in this context corresponds to the amplitude of the EEG signal. The function $p(x)$ denotes the probability density function of the amplitude distribution within a given frequency band. The differential entropy $H(X)$ thus provides a scalar measure of the unpredictability or information content inherent in the EEG signal within that specific frequency band.

This feature extraction process was systematically applied to each frequency band, computing differential entropy values for each individual band. As a result, each EEG sample is represented by a differential entropy feature vector of dimension $N_{cf} = c \times f$, where $c$ denotes the number of EEG channels and $f = 5$ corresponds to the number of frequency bands. This representation effectively captures the spectral-spatial characteristics of the EEG signals, providing a robust and informative foundation for further analysis and interpretation.

\subsubsection{Eye-Movement Feature Representation}
To characterize oculomotor behaviors associated with cognitive and attentional states, a set of representative EM features is extracted from raw eye-tracking signals, covering both low-level physiological variations and high-level behavioral patterns. Based on temporally continuous eye-tracking signals, features are extracted from the following categories. First, pupil-related features are computed along both horizontal and vertical axes. For each axis, the mean, standard deviation, and differential entropy (DE) are calculated within four frequency bands (0–0.2 Hz, 0.2–0.4 Hz, 0.4–0.6 Hz, and 0.6–1 Hz), capturing fine-grained pupil dynamics related to arousal and cognitive load.

Second, dispersion features are extracted to quantify the spatial variability of fixation points, including the mean and standard deviation along both axes, reflecting the spatial stability of eye movements. Third, fixation features are derived in terms of the mean and standard deviation of fixation duration (ms), which are closely associated with attentional engagement and cognitive processing. Fourth, saccade features are computed by extracting the mean and standard deviation of saccade duration (ms) and saccade amplitude (degrees), characterizing rapid eye movements between fixations and visual scanning behaviors. Finally, several aggregate behavioral statistics are considered, including blink frequency, fixation frequency, maximum fixation duration, total and maximum fixation dispersion, saccade frequency, average saccade duration, average saccade amplitude, and average saccade latency, providing a holistic description of EM behavior.

\subsection{Attention-based Feature Modeling}
A key challenge in multimodal neural signal integration lies in the heterogeneity of information representation across modalities. EEG and eye-tracking data differ significantly in temporal structure, spatial granularity, and statistical distribution. Simple concatenation or early fusion may fail to capture complex intra- and inter-modal dependencies. To address this, we propose a hierarchical attention-based framework built upon Transformer-style multi-head attention (MHA), which explicitly models both intra-modality and inter-modality interactions. The framework consists of three core components: modality-specific self-attention for EEG and eye data, and cross-modal attention for guided interaction.

\subsubsection{Self-Attention: Modeling Intra-Modality Dependencies}
Let $f^{\text{eeg}} \in \mathbb{R}^{B \times d_e}$ and $f^{\text{eye}} \in \mathbb{R}^{B \times d_{\text{eye}}}$ denote the modality-specific features extracted by the EEG and eye backbone networks, respectively, where $B$ is the batch size and $d_e$, $d_{\text{eye}}$ are the feature dimensions. We apply self-attention mechanisms to each modality independently to capture contextual dependencies within each modality.

Each self-attention module is constructed using multi-head attention. Given an input $X \in \mathbb{R}^{B \times d}$, we project it into queries $Q$, keys $K$, and values $V$, and split them into $H$ attention heads with head dimension $d_h = d / H$. Each head computes:

\begin{equation}\label{Self-Attention}
	\mathrm{Attention}(Q, K, V) = \mathrm{softmax}\left(\frac{QK^\top}{\sqrt{d_h}}\right)V,
\end{equation}

The outputs of all heads are concatenated and linearly transformed to obtain the final representation. The self-attention operations for each modality are defined as:

\begin{equation}\label{Self-Attention2}
	h^{\text{eeg}} = \mathrm{SelfAttn}_{\text{eeg}}(f^{\text{eeg}}), \quad
	h^{\text{eye}} = \mathrm{SelfAttn}_{\text{eye}}(f^{\text{eye}}),
\end{equation}

where $\mathrm{SelfAttn}_{\text{eeg}}(\cdot)$ and $\mathrm{SelfAttn}_{\text{eye}}(\cdot)$ denote EEG- and eye-specific self-attention modules.

This design enables the model to learn structured dependencies within each modality. For EEG, this includes inter-channel synchronization and cross-frequency coupling; for eye-tracking data, it captures scan path sequences, fixation transitions, and pupil-based state changes.

\subsubsection{Cross-Attention: Modeling Inter-Modality Guidance}
While intra-modality modeling is essential, effective multimodal learning also requires interaction across modalities to exploit their complementary nature. To this end, we introduce a cross-attention mechanism that enables guided representation learning from one modality to another.

Specifically, according to previous studies, the predictive accuracy of emotion recognition based on EEG signals is better than that based on EM signals. Moreover, EEG signals can convey rich information, including temporal and spatial information, while EM signals are susceptible to interference from multiple factors such as illumination and shadow \cite{ZHOU2022108889}. The unique accuracy of EEG signals (which are relatively less controllable) can guide the extraction of EM features\cite{10570465}. We use EEG features as queries and eye features as keys and values, forming a uni-directional cross-attention operation:

\begin{equation}\label{Self-Attention3}
	h^{\text{cross}} = \mathrm{CrossAttn}_{\text{eeg} \leftarrow \text{eye}}(f^{\text{eeg}}, f^{\text{eye}}),
\end{equation}

Here, $\mathrm{CrossAttn}_{\text{eeg} \leftarrow \text{eye}}(\cdot)$ denotes a cross-attention block where EEG queries attend to the eye modality. This guided attention design allows the EEG modality to selectively aggregate semantically relevant information from the eye modality, enhancing the discriminative quality and semantic alignment of the learned representation.

Unlike symmetric dual-attention structures, we adopt a \textit{unidirectional guidance strategy} (EEG $\rightarrow$ Eye) based on the following rationale: EEG signals originate from the central nervous system and possess rich dynamic temporal-spatial patterns, making them more informative and suitable as the dominant modality. In contrast, eye-tracking signals reflect observable behavioral cues such as attention shifts or fatigue levels, which serve as a valuable but supportive source. This directed design reduces noise accumulation from redundant interactions and improves the clarity of semantic fusion.

\subsubsection{Multi-Head Design and Generalization Capacity}
All attention modules adopt a multi-head structure, where each head learns an independent subspace projection and attention distribution, thereby capturing diverse semantic relationships. The final output is computed as:

\begin{equation}\label{Self-Attention4}
	\begin{aligned}
		\mathrm{MHA}(Q, K, V) &= \mathrm{Concat}(\mathrm{head}_1, \dots, \mathrm{head}_H) W^O, \\
		\mathrm{head}_i &= \mathrm{Attention}(Q W^Q_i,\, K W^K_i,\, V W^V_i)
	\end{aligned}
\end{equation}

This structure enhances the model’s representation power and generalization ability by enabling parallel reasoning across multiple attention subspaces. The final representations from self-attention and cross-attention modules are further fused for downstream classification (described in Section~3).

In summary, our hierarchical attention framework jointly captures intra-modal structure and cross-modal semantics through modality-specific self-attention and guided cross-attention, providing a robust and transferable representation for multimodal affective or cognitive state classification.

\subsection{Alignment of Marginal and Conditional Probability Distribution}

\subsubsection{Alignment of Marginal Probability Distribution}
To mitigate the global distribution shift between the source domain $\mathcal{D}_s$ and the target domain $\mathcal{D}_t$, we minimize the Marginal Probability Distribution (MPD) divergence using Maximum Mean Discrepancy (MMD). By mapping data into a Reproducing Kernel Hilbert Space (RKHS) $\mathcal{H}$ via a mapping function $\Phi$, the empirical estimate of MMD is defined as:
\begin{equation}\label{MMD_final1}
	\small
	\begin{aligned}
		\mathcal{L}_{\mathrm{MMD}} &= \left\| \frac{1}{n_s} \sum_{i=1}^{n_s} \Phi(x_{s,i}^{l}) - \frac{1}{n_t} \sum_{j=1}^{n_t} \Phi(x_{t,j}^{u}) \right\|_{\mathcal{H}}^2 \\
		&= \frac{1}{n_s^2} \sum_{i,j=1}^{n_s} k(x_{s,i}, x_{s,j}) + \frac{1}{n_t^2} \sum_{i,j=1}^{n_t} k(x_{t,i}, x_{t,j}) \\
		&\quad - \frac{2}{n_s n_t} \sum_{i=1}^{n_s} \sum_{j=1}^{n_t} k(x_{s,i}, x_{t,j}),
	\end{aligned}
\end{equation}
where $k(x,y) = \langle \Phi(x), \Phi(y) \rangle = e^{-\| x-y \|^2 / \sigma}$ denotes the Gaussian kernel. Minimizing Eq.~(\ref{MMD_final1}) aligns the global statistical properties of the two domains.

\subsubsection{Alignment of Conditional Probability Distribution}
While MMD aligns global distributions, it may neglect class-discriminative boundaries. To address this, we propose Conditional MMD (CMMD) to align the Conditional Probability Distributions (CPD), specifically $P(x_s|y_s=c)$ and $P(x_t|y_t=c)$. Since target labels are unavailable, we utilize pseudo-labels $\hat{y}_t^u = f(x_t^u)$ to approximate the target class-conditional distributions.

We assume that aligning class-wise centroids in RKHS suffices to match the conditional distributions. The CMMD loss is calculated as the average MMD distance between the source and target data within the same category $c \in \{1,\dots,C\}$:
\begin{equation}\label{CMMD_final}
	\small
	\begin{aligned}
		\mathcal{L}_{\mathrm{CMMD}} &= \frac{1}{C}\sum_{c=1}^C \left\| \frac{1}{n_{s}^{c}}\sum_{x_s \in \mathcal{D}_s^c}{\Phi (x_{s})} - \frac{1}{n_{t}^{c}}\sum_{x_t \in \mathcal{D}_t^c}{\Psi (x_{t})} \right\|_{\mathcal{H}}^{2} \\
		&= \frac{1}{C}\sum_{c=1}^C \left[ \frac{1}{(n_{s}^{c})^2}\sum_{i,j \in \mathcal{D}_s^c} k(x_i, x_j) + \frac{1}{(n_{t}^{c})^2}\sum_{i,j \in \mathcal{D}_t^c} k(x_i, x_j) \right. \\
		&\quad \left. - \frac{2}{n_{s}^{c}n_{t}^{c}}\sum_{i \in \mathcal{D}_s^c}\sum_{j \in \mathcal{D}_t^c} k(x_i, x_j) \right],
	\end{aligned}
\end{equation}
where $n_s^c$ and $n_t^c$ denote the number of samples for class $c$ in source and target domains, respectively. As illustrated in Fig.~\ref{fig:method}, the quality of CMMD relies heavily on the accuracy of pseudo-labels, which motivates our proposed confidence-aware weighting strategy.

\subsection{Confidence-Driven Pseudo-Label Weighting}
In the conditional probability distribution alignment process, the lack of ground-truth labels in the target domain necessitates the use of pseudo-labels as surrogate supervision, thereby facilitating conditional distribution alignment between the source and target domains.

To address this issue, we propose a confidence-aware soft pseudo-label weighting mechanism, namely Soft Gaussian Weighting. Unlike traditional methods (e.g., UDA-DDA) that rely on a fixed confidence threshold (e.g., 0.95) to select pseudo-labels, our approach avoids hard filtering and instead adopts a smooth, continuous weighting function based on prediction confidence. While hard thresholding can ensure high precision, it discards a large number of target samples with moderately high confidence, especially during early training. In contrast, Soft Gaussian Weighting allows the model to gradually incorporate unlabeled target samples with different levels of importance, thereby improving data utilization and effectively alleviating the quantity–quality trade-off in pseudo-label learning.

Let $\max(\mathbf{p})$ denote the maximum class probability (i.e., confidence) of a target domain sample. We assume that the distribution of $\max(\mathbf{p})$ approximately follows a truncated Gaussian distribution with mean $\mu_t$ and variance $\sigma_t^2$ at training iteration $t$. Under this assumption, we define the weighting function as:

\begin{equation}
	\lambda(\mathbf{p}) =
	\begin{cases}
		\lambda_{\text{max}} \cdot \exp\left(- \frac{(\max(\mathbf{p}) - \mu_t)^2}{2 \sigma_t^2} \right), & \text{if } \max(\mathbf{p}) < \mu_t \\
		\lambda_{\text{max}}, & \text{otherwise}
	\end{cases}
	\label{eq:gaussian_weighting}
\end{equation}

Here, $\lambda_{\text{max}}$ is the maximum weight (typically 1). Samples with confidence above the mean are assigned full weight, while those below the mean receive an exponentially decaying weight based on their distance from $\mu_t$. This softly filters low-confidence samples rather than removing them entirely.

The parameters $\mu_t$ and $\sigma_t^2$ are not fixed but are dynamically updated during training using exponential moving averages (EMA). Given a mini-batch of target predictions of size $B_U$, we compute:

\begin{align}
	\hat{\mu}_b &= \frac{1}{B_U} \sum_{i=1}^{B_U} \max(\mathbf{p}_i), \\
	\hat{\sigma}_b^2 &= \frac{1}{B_U} \sum_{i=1}^{B_U} (\max(\mathbf{p}_i) - \hat{\mu}_b)^2
\end{align}

Then we update the EMA-based global statistics:

\begin{align}
	\mu_t &= m \cdot \mu_{t-1} + (1 - m) \cdot \hat{\mu}_b, \\
	\sigma_t^2 &= m \cdot \sigma_{t-1}^2 + (1 - m) \cdot \frac{B_U}{B_U - 1} \cdot \hat{\sigma}_b^2
\end{align}

where $m$ is the momentum factor (e.g., $0.999$), and initial values are set to $\mu_0 = \frac{1}{C}$ and $\sigma_0^2 = 1.0$.

A desirable property of this approach is that it ensures at least a lower bound on the effective utilization of target domain data. Let $f(\mathbf{p})$ denote the expected sample weight across the target domain:

\begin{equation}
	f(\mathbf{p}) = \mathbb{E}_{\mathcal{D}_U}[\lambda(\mathbf{p})]
\end{equation}

We can derive a theoretical lower bound:

\begin{equation}
	f(\mathbf{p}) \in \left[ \frac{\lambda_{\text{max}}}{2} \left( 1 + \exp\left(- \frac{(1/C - \mu_t)^2}{2\sigma_t^2} \right) \right), \; \lambda_{\text{max}} \right]
\end{equation}

This guarantees that at least approximately half of the pseudo-labeled target samples contribute to training loss with non-zero weight at all times.

As training progresses, the model becomes more confident, resulting in:
\begin{itemize}
	\item $\mu_t$ gradually increases;
	\item $\sigma_t$ gradually decreases.
\end{itemize}

\begin{figure}[t]
	\centering
	\includegraphics[width=\linewidth]{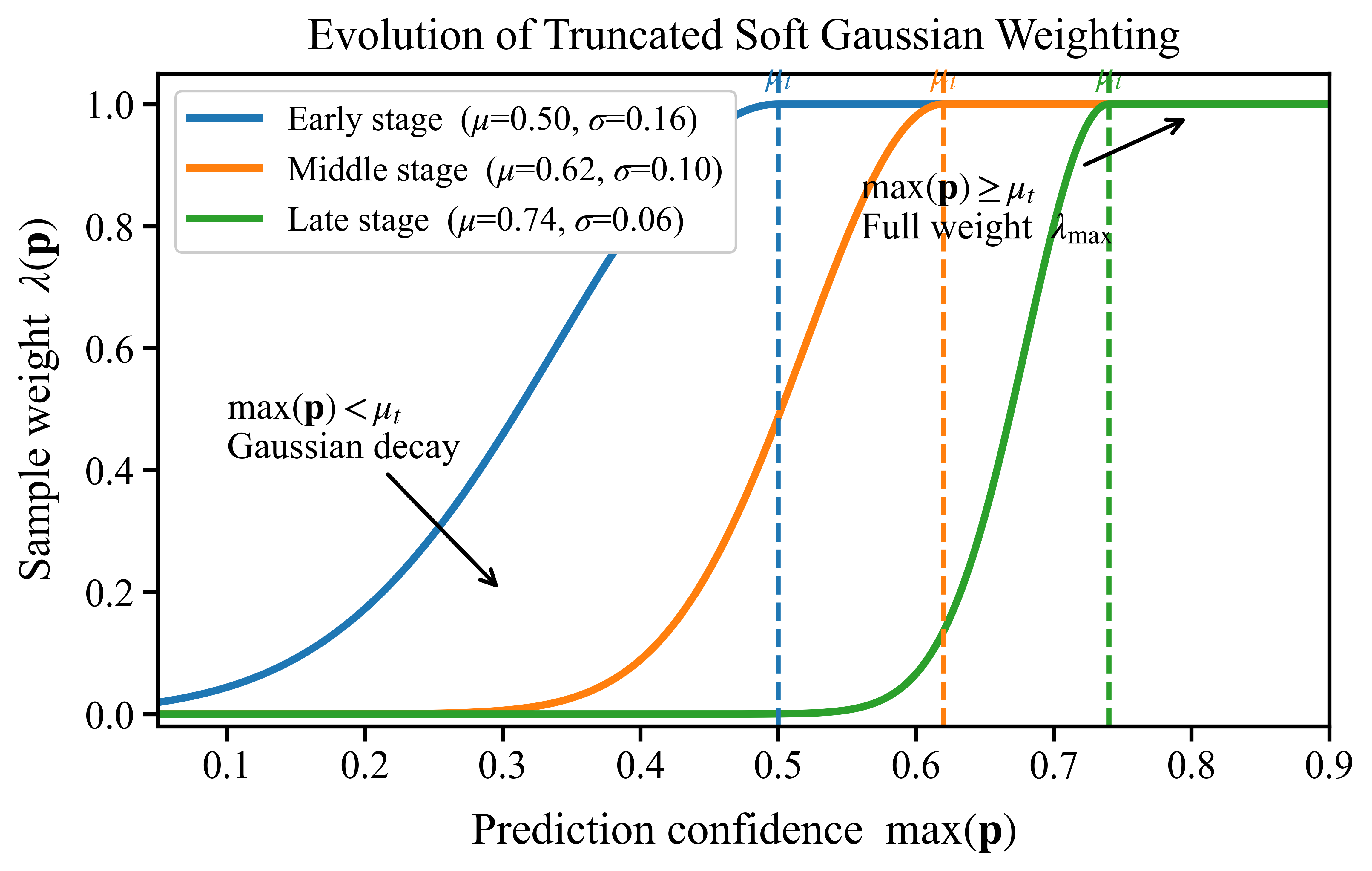}
	\caption{Evolution of the truncated soft Gaussian weighting function.
		Samples with prediction confidence above the adaptive threshold $\mu_t$ receive the full weight $\lambda_{\max}$, while those below $\mu_t$ are softly down-weighted following a Gaussian decay.}
	\label{fig:soft-gaussian-weighting}
\end{figure}

As illustrated in Fig.~\ref{fig:soft-gaussian-weighting}, this evolution makes the truncated Gaussian weighting function increasingly steeper, thereby further suppressing low-confidence predictions. Consequently, the model smoothly transitions from ``quantity-emphasis'' to ``quality-emphasis'' in pseudo-label selection, leading to more stable training and improved generalization.

\subsection{Uniform Alignment for Class-Wise Pseudo-Label Balancing}
In cross-domain emotion recognition tasks, pseudo-labels in the target domain often suffer from severe class imbalance. This imbalance stems from two main sources: (1) the model's varying classification performance across classes, which favors “easier” classes while neglecting “harder” ones; (2) in early training stages, many low-confidence predictions are filtered out, further skewing the class distribution. This phenomenon not only affects the model's ability to learn from tail classes but also degrades the effectiveness of CPD alignment, especially when class-conditional statistics are estimated using a limited set of biased pseudo-labels.

To address this issue, we introduce UA. This module adjusts the predicted probability distributions of pseudo-labels in the target domain to reduce class imbalance, while preserving the semantic information encoded in the predictions.

In class-conditional alignment methods such as CMMD, alignment is performed separately for each class. However, if the pseudo-labels in the target domain are highly imbalanced—e.g., some classes have very few confident predictions—the estimation of class-wise kernel mean embeddings becomes unreliable. This compromises the stability and fairness of the alignment loss. Therefore, we aim to softly balance the class-wise contributions of pseudo-labels to ensure equitable alignment across all categories.

The goal of UA is to encourage the pseudo-label distributions to be closer to a uniform distribution across classes. Specifically, let $\hat{p}(y)$ denote the current batch-wise predicted class distribution over the target domain:

\begin{equation}
	\hat{p}(y) = \frac{1}{B_U} \sum_{i=1}^{B_U} p(y|x_i^t)
\end{equation}

We compare $\hat{p}(y)$ to the ideal uniform distribution $u(y) = \frac{1}{C}$, and compute a scaling factor for each class. Then, for each target sample $p_i = p(y|x_i^t)$, we perform the following normalization:

\begin{equation}
	p_i^{\text{aligned}} = \frac{p_i \cdot \left( \frac{u(y)}{\hat{p}(y)} \right)^\tau }{\sum_{j} p_{i,j} \cdot \left( \frac{u_j}{\hat{p}_j} \right)^\tau }
	\label{eq:ua_adjust}
\end{equation}

Here, $\tau$ is a temperature coefficient that controls the adjustment strength. This normalization suppresses the contributions of over-represented classes and amplifies those of under-represented classes, thereby encouraging more balanced class-wise usage of pseudo-labels.

To avoid over-correction in early training when predictions are still unstable, we introduce a dynamic scheduling factor $\alpha_t$ using a sigmoid decay function:

\begin{equation}
	\alpha_t = \frac{\alpha_0}{1 + \exp\left( \frac{t - T_0}{k} \right)}
	\label{eq:alpha_schedule}
\end{equation}

Here, $\alpha_0$ is the initial alignment strength, $T_0$ is the epoch at which the decay inflection occurs, and $k$ controls the decay slope. We then interpolate between the aligned and original distributions:

\begin{equation}
	\tilde{p}(y|x) = \alpha_t \cdot u(y) + (1 - \alpha_t) \cdot \hat{p}(y|x)
\end{equation}

This ensures that strong alignment only occurs after the model becomes reasonably calibrated.

The output of the UA module is a set of class-rebalanced soft pseudo-labels $\tilde{p}(y|x_t^u)$. These adjusted probabilities are used in the subsequent CMMD loss to compute class-wise kernel mean embeddings with improved fairness and reliability.

By introducing UA, we ensure that each class contributes meaningfully to the conditional alignment process, even when pseudo-labels are sparse or uncertain. When combined with confidence-aware weighting from the Soft Gaussian module, this provides both instance-level and class-level robustness to noisy pseudo-labels.
\subsection{Training Procedure}
Our proposed framework is trained in an end-to-end manner by jointly optimizing classification and distribution alignment objectives. In each mini-batch, labeled source domain samples $(x_s^l, y_s^l)$ and unlabeled target domain samples $x_t^u$ are first processed by modality-specific backbone networks for EEG and eye-movement signals, followed by self-attention modules to enhance intra-modality dependencies. The resulting features are fused through a cross-attention mechanism, where EEG serves as the query and eye features provide keys and values, producing a multimodal representation that is fed into a shared classification head to generate class probability predictions. For target domain samples, initial pseudo-labels $\mathbf{p}(y|x_t^u)$ are obtained from the classifier outputs and refined through Soft Gaussian Weighting, which assigns confidence-based weights $\lambda$ according to Eq.~(\ref{eq:gaussian_weighting}), and UA, which adjusts pseudo-label distributions via Eq.~(\ref{eq:ua_adjust}) to balance class representation, resulting in class-balanced soft labels $\tilde{p}(y|x_t^u)$. The overall training loss is defined as 
\begin{equation}
	\mathcal{L} = \mathcal{L}_{\mathrm{cls}} + \gamma_{\mathrm{mmd}} \mathcal{L}_{\mathrm{MMD}} + \gamma_{\mathrm{cmmd}} \mathcal{L}_{\mathrm{CMMD}},
\end{equation}
where $\mathcal{L}_{\mathrm{cls}}$ is the cross-entropy loss on labeled source samples, $\mathcal{L}_{\mathrm{MMD}}$ is the marginal distribution alignment term in Eq.~(\ref{MMD_final1}), and $\mathcal{L}_{\mathrm{CMMD}}$ is the conditional alignment loss computed using soft pseudo-labels and confidence weights, with $\gamma_{\mathrm{mmd}}$ and $\gamma_{\mathrm{cmmd}}$ controlling their relative importance. Gradients of $\mathcal{L}$ are back-propagated through the entire network, and parameters are updated via the Adam optimizer. The statistics $\mu_t$ and $\sigma_t^2$ for Soft Gaussian Weighting, as well as $\alpha_t$ for UA, are updated at each iteration using exponential moving averages and a predefined scheduling strategy. This process is repeated until convergence, during which the model gradually shifts from quantity-emphasis to quality-emphasis in pseudo-label utilization, while maintaining balanced class-wise contributions for stable and robust cross-domain adaptation.

\begin{algorithm}[!h]
	\caption{: Training Algorithm of the Proposed Framework}\label{alg:Proposed}
	\renewcommand{\algorithmicrequire}{\textbf{Input:}}
	\renewcommand{\algorithmicensure}{\textbf{Output:}}
	\begin{algorithmic}[1]
		\REQUIRE $D_s = \{x_s^l, y_s^l\}_{i=1}^{N_s}$, \quad $D_t = \{x_t^u\}_{j=1}^{N_t}$
		\ENSURE Predicted labels for the target domain
		\STATE Randomly sample a mini-batch $D_{batch}^s$ and $D_{batch}^t$ from $D_s$ and $D_t$ respectively;
		\STATE Extract modality-specific EEG and eye-movement features via backbone networks, then apply self-attention modules for intra-modality enhancement;
		\STATE Fuse the attended EEG and eye features via cross-attention (EEG as query, eye as key/value) to obtain multimodal representation;
		\STATE Generate initial pseudo-labels $\mathbf{p}(y|x_t^u)$ for $x_t^u \in D_{batch}^t$ from the classifier;
		\STATE Apply Soft Gaussian Weighting (Eq.~\ref{eq:gaussian_weighting}) to assign confidence weights $\lambda$, and perform UA (Eq.~\ref{eq:ua_adjust}) to obtain class-balanced soft labels $\tilde{p}(y|x_t^u)$;
		\STATE Compute $\mathcal{L}_{\mathrm{cls}}$, $\mathcal{L}_{\mathrm{MMD}}$ (Eq.~\ref{MMD_final1}), and $\mathcal{L}_{\mathrm{CMMD}}$ using weighted soft pseudo-labels; 
		\STATE Update network parameters via Adam optimizer to minimize $\mathcal{L} = \mathcal{L}_{\mathrm{cls}} + \gamma_{\mathrm{mmd}} \mathcal{L}_{\mathrm{MMD}} + \gamma_{\mathrm{cmmd}} \mathcal{L}_{\mathrm{CMMD}}$, while updating $\mu_t$, $\sigma_t^2$, and $\alpha_t$ with EMA scheduling;
		\RETURN Predicted labels for $D_t$.
	\end{algorithmic}
\end{algorithm}
\begin{table*}[h]
	\centering
	\caption{The experiment results of different methods on the task of cross-subject emotion recognition (\%)}
	\label{tab:results}
	\tiny
	\renewcommand{\arraystretch}{0.85} 
	\setlength{\aboverulesep}{0.3ex}    
	\setlength{\belowrulesep}{0.3ex}    
	\resizebox{\textwidth}{!}{%
		\begin{tabular}{lcccccc}
			\toprule
			Methods & \multicolumn{3}{c}{SEED} & \multicolumn{3}{c}{SEED-IV} \\
			\cmidrule(lr){2-4} \cmidrule(lr){5-7} 
			& Acc & Macro-F1 & AUC & Acc & Macro-F1 & AUC \\
			\midrule
			DGCNN\cite{song2018eeg} & 79.95$\pm$9.02 & - & - & - & - & - \\
			MHESA\cite{ZHU2024124001} & - & - & - & 83.25$\pm$9.98 & - & - \\
			CFDA-CSF\cite{article} & 90.04$\pm$5.46 & - & - & 89.60$\pm$6.65 & - & - \\
			MMDA\cite{10819285} & 94.82$\pm$2.41 & 94.75$\pm$2.47 & - & 85.54$\pm$8.11 & 85.28$\pm$8.05 & - \\
			CMSLNet\cite{10575932} & - & - & - & 83.15$\pm$9.84 & - & - \\
			MACDB\cite{JIMENEZGUARNEROS2025113238} & 90.49$\pm$4.04 & - & - & 83.02$\pm$4.67 & - & - \\
			CSMM\cite{10938180} & 94.96$\pm$5.27 & 95.21$\pm$7.96 & 96.20$\pm$3.98 & 89.82$\pm$6.22 & 90.03$\pm$6.19 & 93.01$\pm$4.25 \\
			\textbf{Ours} & \textbf{94.68$\pm$3.91} & \textbf{94.69$\pm$3.74} & \textbf{97.68$\pm$2.50} & \textbf{92.00$\pm$5.29} & \textbf{92.88$\pm$4.64} & \textbf{92.02$\pm$5.05} \\
			\bottomrule
	\end{tabular}}
\end{table*}

\section{Experiments and Results}\label{sec:experimentresults}
\subsection{Emotion Datasets}
The primary datasets used in this study are SEED and SEED-IV, which are publicly available from the Brain-like Computing and Machine Intelligence Center at Shanghai Jiao Tong University. The SEED dataset comprises experiments involving fifteen Chinese participants. Each participant completed three sessions, with each session containing fifteen trials. During the trials, participants viewed Chinese film clips designed to elicit positive, neutral, and negative emotions \cite{zheng2015investigating} . The SEED-IV dataset also involved 15 subjects across three sessions conducted on separate days; each session included 24 trials. In this dataset, participants watched film clips intended to evoke happiness, sadness, neutrality, or fear. For both SEED and SEED-IV, EEG data were recorded using a 62-channel ESI NeuroScan System, EM data weresimultaneously recorded using SMI ETG eye-tracking glasses\cite{zheng2015investigating,8283814} . 

The DEAP dataset includes data from 32 participants (16 male, 16 female), who watched 40 one-minute music videos while their EEG signals were recorded via 32 electrodes. After each video, participants rated their levels of arousal, valence, dominance, and liking using Self-Assessment Manikins on a continuous scale from 1 to 9\cite{koelstra2011deap}.

\subsection{Implementation Details and Experimental Setup}
The proposed HADUA framework is implemented in PyTorch 2.1.0 and trained end-to-end using the Adam optimizer. The initial learning rate is set to $1\times 10^{-4}$, with a weight decay of $5\times 10^{-5}$, and a batch size of 64. Training proceeds for up to 100 epochs, with early stopping triggered if the validation accuracy shows no improvement over 15 consecutive epochs. EEG and eye-movement signals are processed by separate modality-specific backbone networks. EEG features are extracted using a two-layer bidirectional GRU, while eye-movement features are obtained via a one-layer CNN followed by a GRU. Each feature stream is subsequently enhanced by a self-attention module. Cross-attention fusion is then applied, with EEG features serving as queries and eye features providing keys and values. The resulting fused representation is passed through a two-layer fully connected classification head with ReLU activation. The overall loss function combines the source-domain classification loss, the marginal MMD loss, and the conditional MMD loss, with $\gamma_{\mathrm{mmd}} = 0.5$ and $\gamma_{\mathrm{cmmd}} = 0.5$. For pseudo-label refinement, Soft Gaussian Weighting dynamically adjusts sample-wise confidence using an exponential moving average of $\sigma_t$ with a momentum of 0.9. Concurrently, UA applies a scheduling parameter $\alpha_t$ that linearly increases from 0 to 1 over the course of training. All experiments were conducted on an NVIDIA RTX 3090 GPU. The source code will be made publicly available upon publication.

\subsection{Result}
Table~\ref{tab:results} summarizes the cross-subject emotion recognition performance on SEED and SEED-IV under a consistent evaluation protocol, with \emph{Acc}, \emph{Macro-F1}, and \emph{AUC} serving as the primary evaluation metrics. We compare HADUA against representative EEG/EM multimodal baselines and recent cross-subject adaptation methods, including DGCNN~\cite{song2018eeg}, MHESA~\cite{ZHU2024124001}, CFDA-CSF~\cite{article}, MMDA~\cite{10819285}, CMSLNet~\cite{10575932}, MACDB~\cite{JIMENEZGUARNEROS2025113238}, and CSMM~\cite{10938180}. On SEED, HADUA achieves an accuracy of 94.68\% Acc and a Macro-F1 score of 94.69\%, demonstrating performance highly competitive with the strongest multimodal adaptation baselines, such as MMDA~\cite{10819285} and CSMM~\cite{10938180}. Notably, HADUA obtains the highest AUC of 97.68\%, indicating more reliable ranking and decision confidence under cross-subject distribution shifts. This suggests that the proposed fusion-and-adaptation pipeline improves not only point-wise classification accuracy but also the overall discriminative quality of the learned representations. On the more challenging SEED-IV dataset, HADUA shows clear advantages, reaching 92.00\% accuracy and a 92.88\% Macro-F1 score, thereby outperforming prior methods including CFDA-CSF~\cite{article} and CSMM~\cite{10938180}. The performance gains observed on SEED-IV further validate the effectiveness of HADUA in multi-class settings, where issues such as pseudo-label noise and class-wise bias are often amplified during cross-subject adaptation. In summary, the results demonstrate that HADUA delivers robust and consistent improvements across datasets, affirming its suitability for practical cross-subject multimodal emotion recognition applications.
\section{Discussion}
\subsection{Confusion Matrices}
On the SEED dataset, the proposed method achieves consistently high recognition accuracy across all emotion categories, with accuracies of 92.92\% for negative, 94.30\% for neutral, and 96.73\% for positive emotions. Errors are primarily concentrated between negative and neutral categories, reflecting their proximity in the emotional valence dimension, whereas positive emotion exhibits more stable and separable neural–ocular patterns. On the SEED-IV dataset, robust class-wise performance is also observed, with recognition accuracies of 92.60\% for neutral, 88.83\% for sad, 97.40\% for fear, and 90.50\% for happy emotions. Misclassifications mainly occur among affectively adjacent categories, particularly between emotions with similar arousal or valence characteristics, such as sad and neutral, as well as happy and fear. Notably, fear emerges as the most discriminative emotion category, indicating strong and distinctive multimodal correlates.

The results on both datasets demonstrate that the proposed framework achieves not only high recognition accuracy but also balanced performance across emotion categories. The observed errors are largely confined to semantically and affectively neighboring emotions, highlighting the robustness and interpretability of the model under cross-subject conditions.
\begin{figure}[t]
	\centering
	\begin{minipage}{0.49\linewidth}
		\centering
		\includegraphics[width=0.9\linewidth]{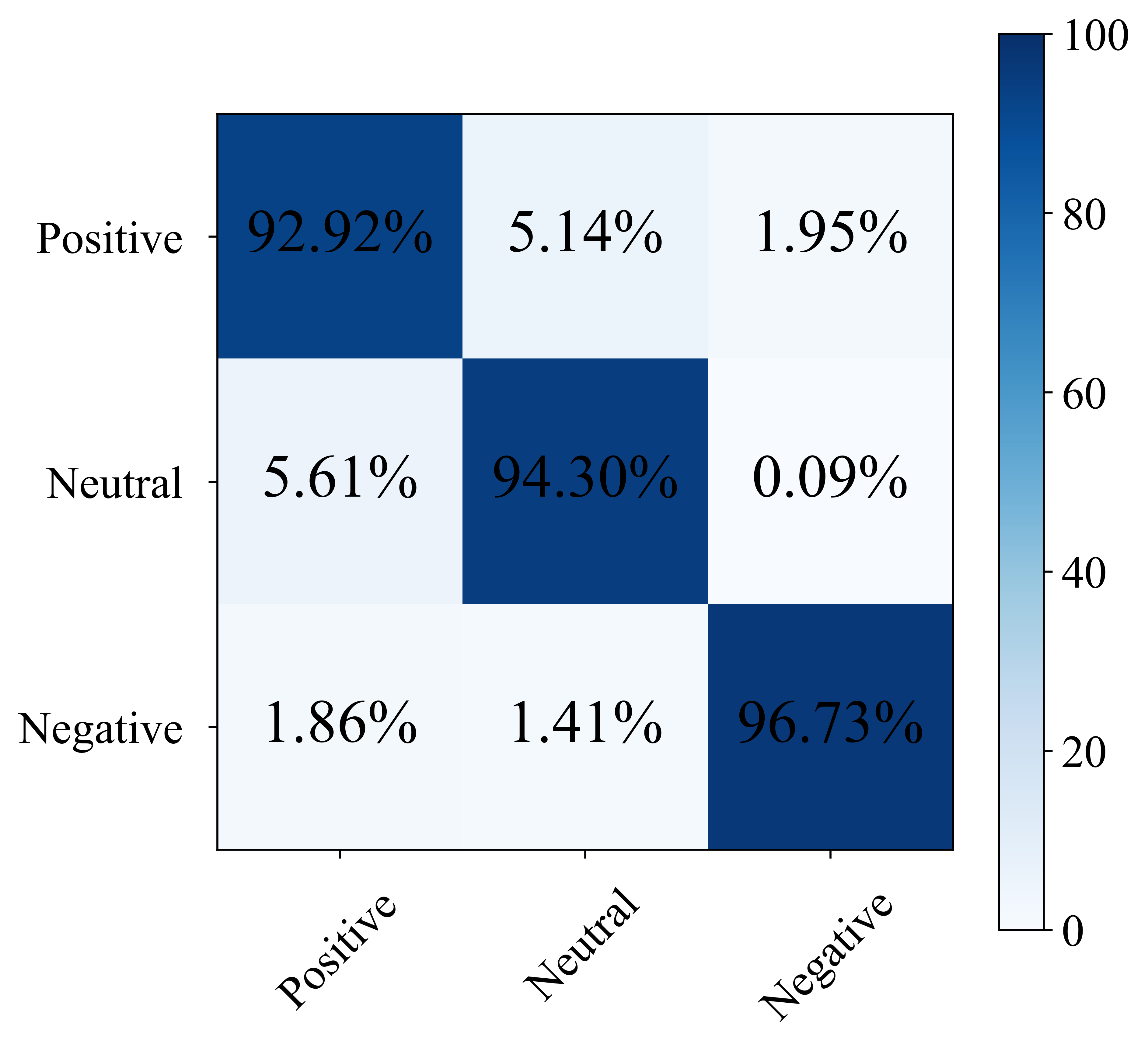}
		\label{fg:RGNN}
		\vspace{2pt}
		\linebreak (a) HADUA on SEED
	\end{minipage}
	\begin{minipage}{0.49\linewidth}
		\centering
		\includegraphics[width=0.9\linewidth]{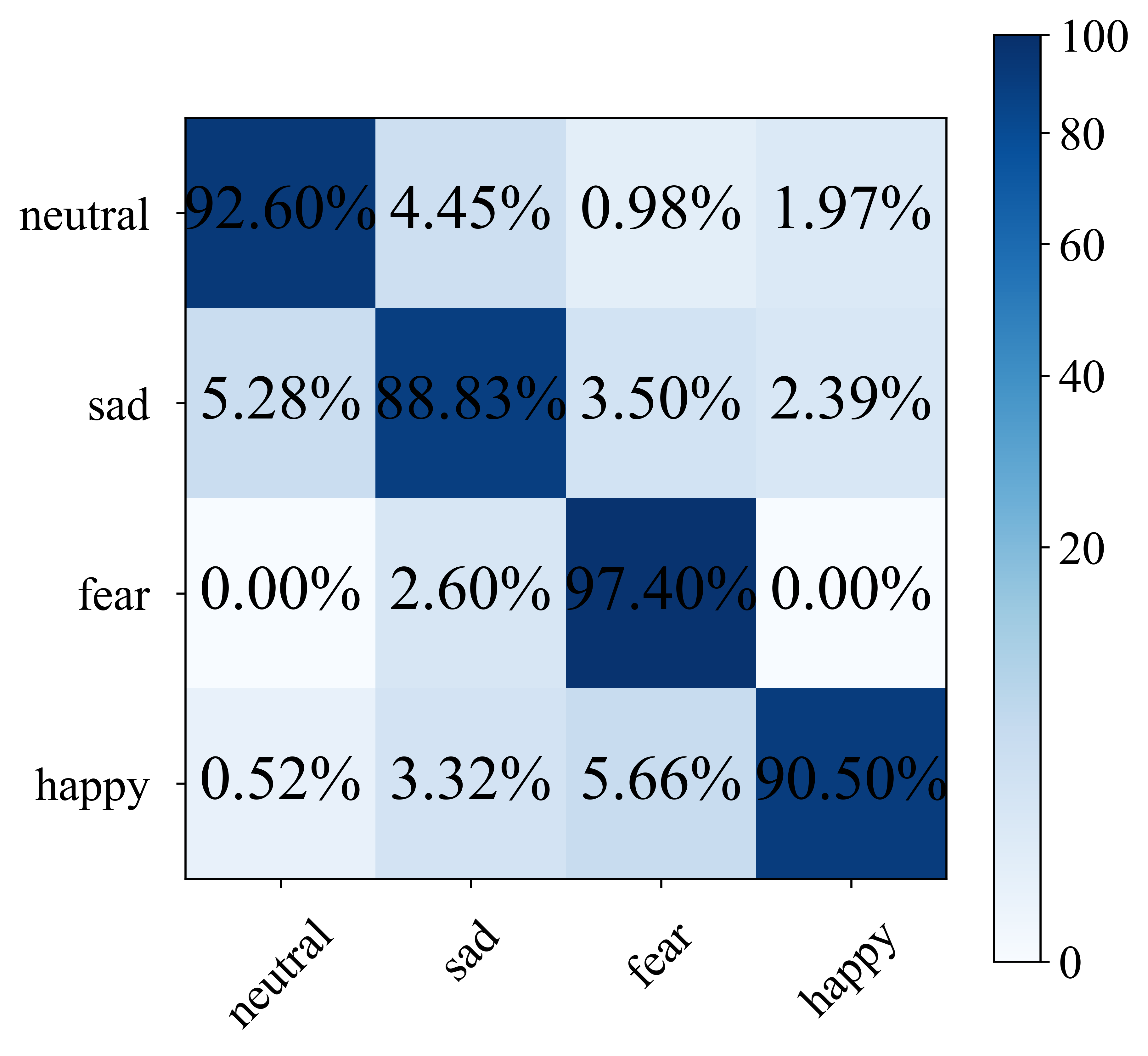}
		\label{fg:JTSR}
		\linebreak (a) HADUA on SEED-IV
	\end{minipage}
	\caption{Confusion matrices of HADUA on (a) SEED and (b) SEED-IV datasets. For SEED, the emotion categories are Negative, Neutral, and Positive. For SEED-IV, the categories are neutral, sad, fear, and happy.}
	\label{fg:ConfusionMatrix}
\end{figure}

\begin{table}[htbp]
	\centering
	\caption{Standard deviation of per-class accuracies on SEED (3-class) and SEED-IV (4-class).}
	\label{tab:std}
	\small
	\renewcommand{\arraystretch}{0.95}   
	\setlength{\tabcolsep}{10pt}          
	\setlength{\aboverulesep}{0.4ex}
	\setlength{\belowrulesep}{0.4ex}
	\begin{tabular}{lcc}
		\toprule
		Method & SEED (3-class) & SEED-IV (4-class) \\
		\midrule
		CSMM\cite{10938180} & 4.17 & -- \\
		MHESA\cite{ZHU2024124001} & -- & 3.64 \\
		MMDA\cite{10819285} & 3.48 & 5.01 \\
		MACDB\cite{JIMENEZGUARNEROS2025113238} & 3.55 & 8.10 \\
		CFDA-CSF\cite{article} & 2.45 & 3.41 \\
		Att-LMF\cite{chen2024comprehensive} & -- & 4.13 \\
		\textbf{HADUA} & \textbf{1.92} & \textbf{3.21} \\
		\bottomrule
	\end{tabular}
\end{table}

We additionally examine the inter-class balance of recognition by analyzing the standard deviation of per-class accuracies, as shown in Table~\ref{tab:std}. This statistic offers a complementary perspective to average accuracy: while high accuracy reflects strong global performance, a low standard deviation indicates that the classifier does not rely disproportionately on a subset of categories. HADUA achieves the lowest deviations across both datasets (1.92 on SEED and 3.21 on SEED-IV), substantially outperforming all competing methods. This advantage primarily stems from the proposed UA mechanism, which dynamically regulates pseudo-label distributions to alleviate class imbalance and ensure fair contribution from all emotion categories. As a result, HADUA captures emotion-discriminative features that generalize across both dominant and subtle classes, thereby mitigating the intrinsic difficulty of differentiating semantically similar emotions (e.g., \emph{Sad} vs. \emph{Neutral}). These findings confirm that HADUA delivers not only state-of-the-art recognition rates but also balanced cross-class consistency, an essential property for building trustworthy affective recognition systems. 
\begin{table*}[t]
	\centering
	\caption{Impact of Each Component as Increasingly Added to HADUA}
	\label{tab:hadua-ablation}
	\normalsize
	\begin{tabular}{lcc}
		\toprule
		Method & SEED & SEED-IV \\
		\midrule
		Deep Feedforward Network & 85.18$\pm$05.60 & 81.00$\pm$11.11 \\
		+ Hierarchical Attention-based Multimodal Fusion  & 90.94$\pm$04.62 & 88.94$\pm$06.89\\
		+ Marginal distribution alignment & 93.50$\pm$03.99 & 90.89$\pm$07.07  \\
		+ Dynamic Gaussian Confidence-weighted Domain Adaptation & 94.42$\pm$04.22 & 91.46$\pm$07.11 \\
		+ UA Mechanism (HADUA)  & 94.68$\pm$03.91 &  92.00$\pm$05.29 \\ 
		\bottomrule
	\end{tabular}
	\vspace{1mm}
\end{table*}

\subsection{Ablation Results}
To examine the contribution of each component in the proposed framework, we conduct ablation studies by progressively incorporating different modules, with results summarized in Table~\ref{tab:hadua-ablation}. Starting from a deep feedforward network baseline, the model achieves moderate performance, with classification accuracies of 85.18\% on SEED and 81.00\% on SEED-IV, indicating that simple feature concatenation is insufficient to capture the complex dependencies inherent in multimodal physiological signals. 
Introducing the hierarchical attention-based multimodal fusion module leads to a substantial performance improvement, boosting the classification accuracy to 90.94\% on SEED and 88.94\% on SEED-IV. This gain highlights the effectiveness of self-attention in enhancing modality-specific representations, as well as cross-attention in exploiting complementary information between EEG and EM modalities. 
When marginal distribution alignment is further incorporated, the performance consistently improves to 93.50\% and 90.89\% on SEED and SEED-IV, respectively, demonstrating the importance of mitigating global distribution discrepancies under cross-subject settings. 
The addition of the dynamic Gaussian confidence-weighted domain adaptation module yields further gains, raising the classification accuracy to 94.42\% on SEED and 91.46\% on SEED-IV, which confirms its role in alleviating the quantity–quality trade-off in pseudo-label learning and stabilizing the adaptation process. 
Finally, the full model equipped with the Uniform Alignment (UA) mechanism achieves the best performance, reaching 94.68\% on SEED and 92.00\% on SEED-IV. Overall, the ablation results reveal a clear and consistent upward trend as more components are integrated, indicating strong complementarity among the proposed modules and validating the effectiveness of the complete HADUA framework for robust cross-subject emotion recognition.
\subsection{Representation Visualization}
\subsubsection{t-SNE Visualization of Feature Learning Dynamics}
To better understand the progressive learning dynamics and domain adaptation capability of our HADUA framework, we employ t-SNE to visualize the evolution of multimodal fused features across training stages. The visualization utilizes key parameters: a perplexity of 30, PCA initialization, and 2000 iterations for optimal convergence.

\begin{figure}[t]
	\centering
	\includegraphics[width=\linewidth]{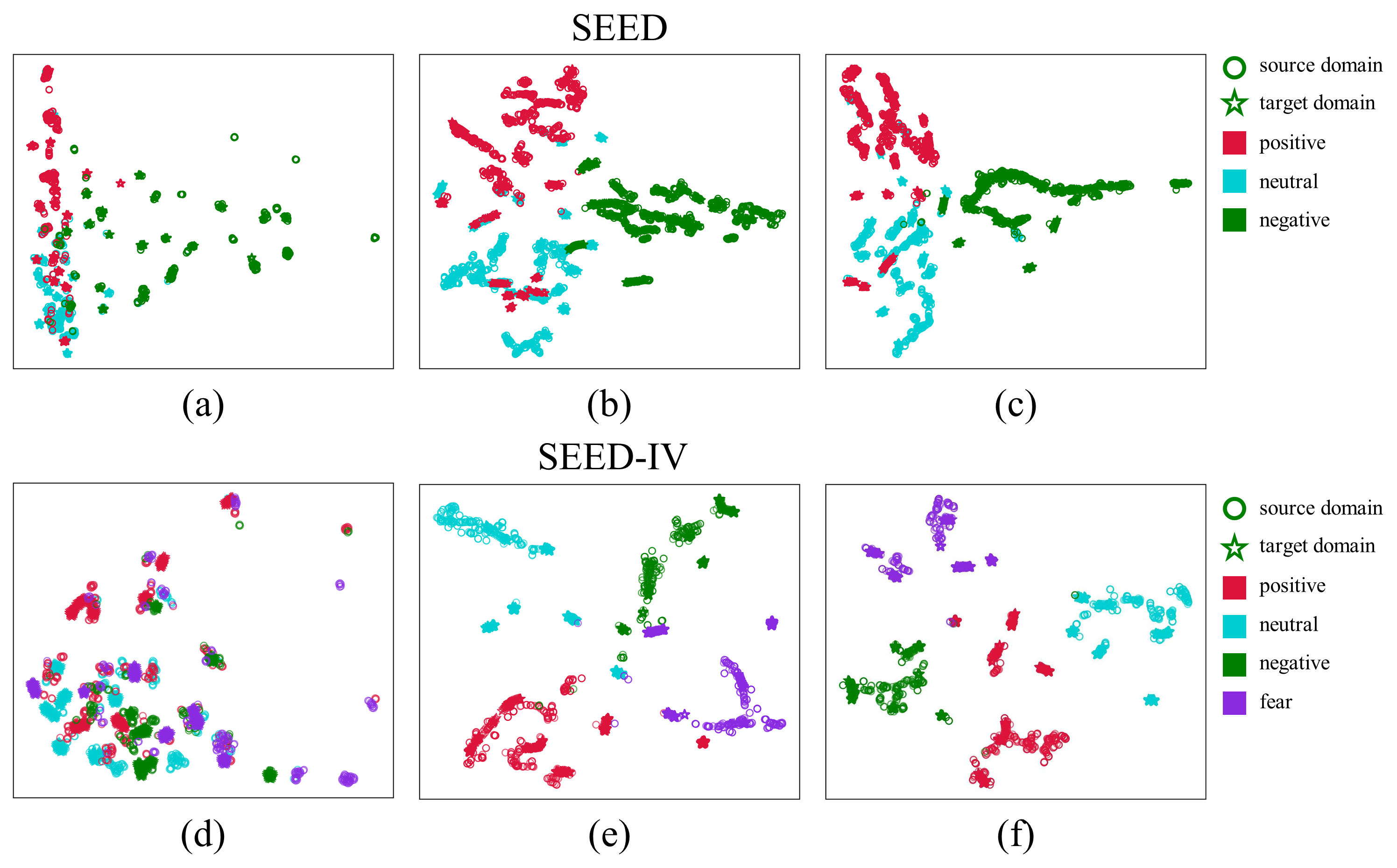}
	\caption{Visualization of feature evolution on the SEED and SEED-IV datasets. (a–c) show the feature distributions of SEED at epochs 0, 50, and 200, respectively; (d–f) present the feature distributions of SEED-IV at the corresponding epochs. Different colors denote different emotion categories, while circular and star markers represent source-domain and target-domain samples, respectively.}
	\label{fig:tsne}
\end{figure}

Based on t-SNE visualizations for one subject from the SEED and SEED-IV datasets, the progressive evolution of feature distributions across epochs (0, 50, 200) effectively demonstrates both the learning dynamics and domain alignment capability of HADUA. As illustrated in Fig.~\ref{fig:tsne}, for SEED (first row), three emotional categories are visualized with distinct colors, where circular markers denote source domain samples and star markers represent target domain samples. At epoch 0 (a), features from both domains exhibit substantial overlap with no clear alignment. By epoch 50 (b), distributions begin to form discernible emotion clusters while simultaneously showing improved alignment between source and target domains within each category. At epoch 200 (c), each emotion forms compact, well-separated clusters with clear decision boundaries, where source and target samples demonstrate remarkable alignment within each cluster.Similarly, for SEED-IV (second row) containing four emotional categories, the same progressive enhancement pattern is observed. Initially confused distributions at epoch 0 (d) gradually evolve into distinct, tightly clustered formations by epoch 200 (f), with source and target samples converging within each category. The significantly reduced intra-category scatter and improved inter-class separation are accompanied by effective domain invariance, as evidenced by the intermingling of source (circles) and target (stars) samples within the same emotion clusters. The observed cluster formation and domain alignment across both datasets provide compelling visual evidence for the framework's ability to learn discriminative and domain-invariant representations.

\subsubsection{Topographic Analysis of Feature Importance}
\begin{figure}[t]
	\centering
	\includegraphics[width=\linewidth]{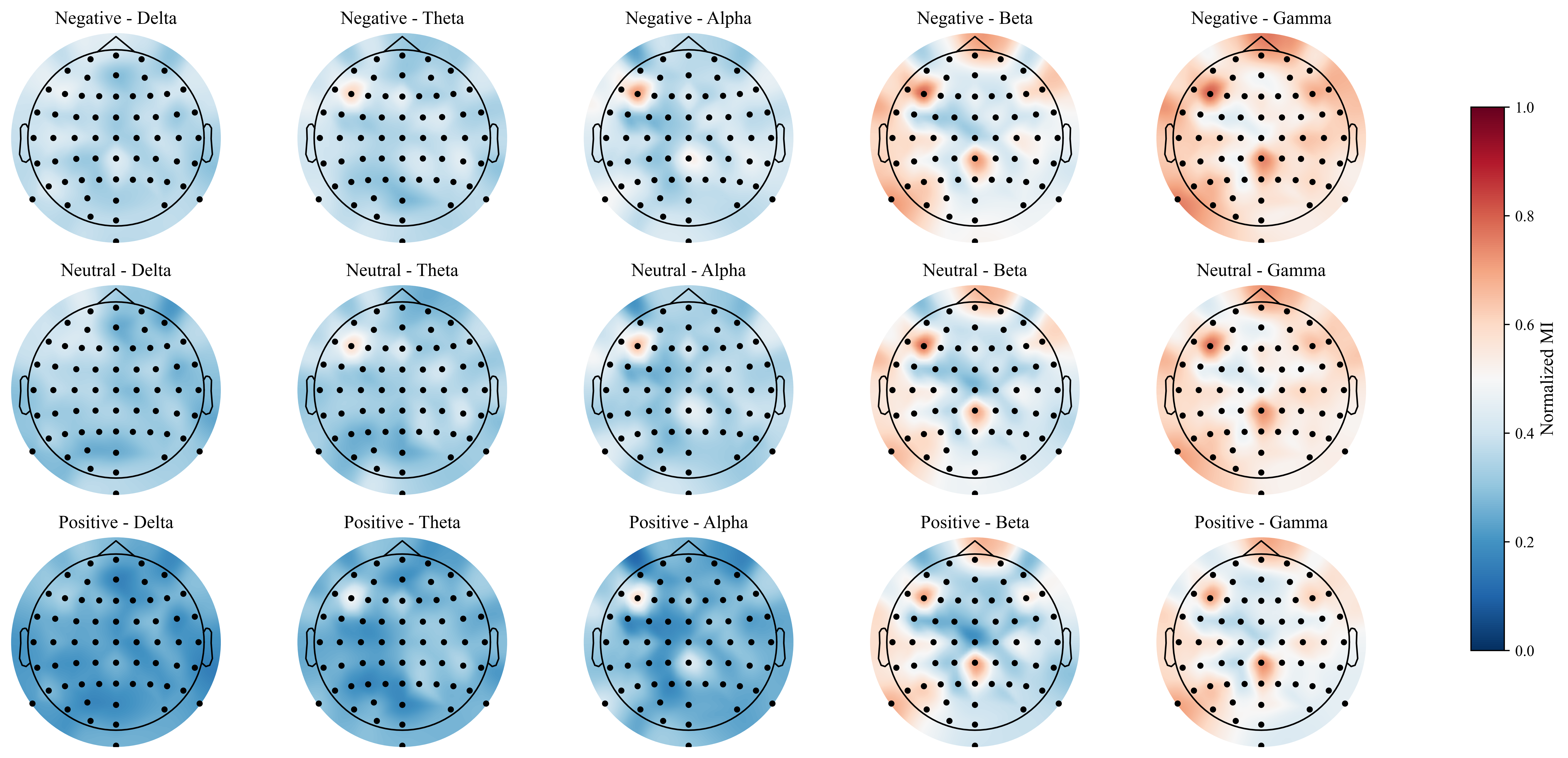}
	\caption{Feature importance (Mutual Information) on SEED. Frontal electrodes (F5, Fpz) exhibit the highest discriminative power, peaking in the Gamma band.}
	\label{fig:seedaverage}
\end{figure}

\begin{figure}[t]
	\centering
	\includegraphics[width=\linewidth]{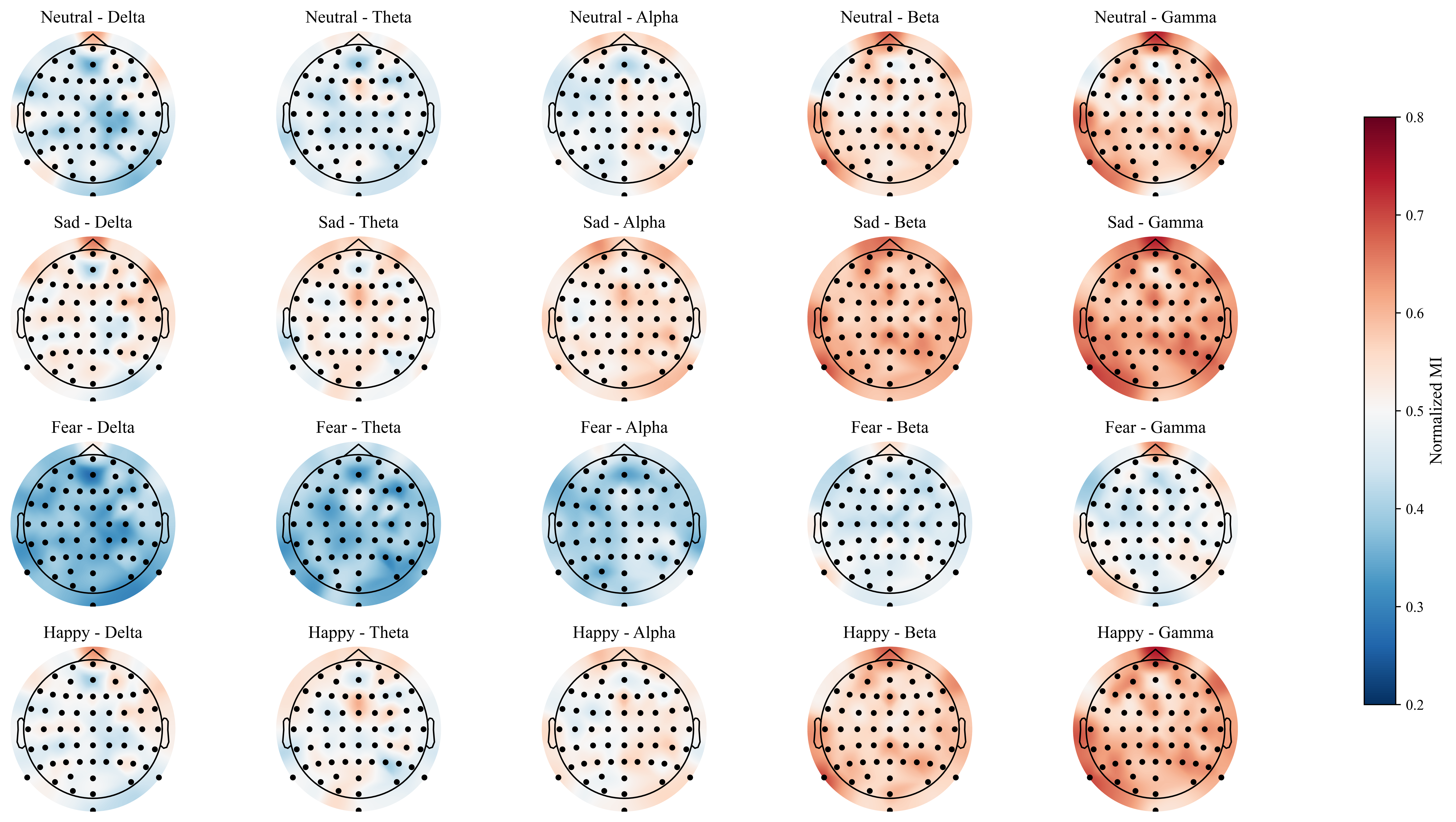}
	\caption{Feature importance (Mutual Information) on SEED-IV. Posterior regions (PO7) and frontal areas show prominent involvement, while Fear exhibits consistently lower values.}
	\label{fig:seedivaverage}
\end{figure}

To identify the neurophysiological biomarkers that contribute most significantly to emotion recognition and validate the physiological plausibility of our HADUA framework, we computed mutual information (MI) between the input EEG differential entropy features (62 channels × 5 frequency bands) and the model's emotion prediction probabilities. This analysis reveals which original EEG characteristics are most informative for the model's decision-making process. Topographic maps were generated using a 62-channel EEG systems for both SEED and SEED-IV datasets, with MI values averaged across all subjects to obtain robust patterns. As illustrated in Fig.~\ref{fig:seedaverage}, our analysis revealed distinct spatial patterns of feature importance across emotional categories. In the SEED dataset, frontal regions demonstrated predominant discriminative power, with F5 (0.6249), Fpz (0.5146), and F7 (0.4964) emerging as the most important channels. This frontal dominance was particularly pronounced for negative emotions, where F5 consistently ranked first across all frequency bands, reaching maximum MI values in gamma band (0.7940).In the SEED-IV dataset, we observed a more distributed pattern with increased importance of posterior regions. As shown in Fig.~\ref{fig:seedivaverage}, the overall top channels included PO7 (0.5516), Fp2 (0.5486), and F2 (0.5477). Notably, PO7 demonstrated exceptional discriminative power across all four emotions, ranking first in both neutral (0.5671) and happy (0.5854) emotions when averaged across frequency bands.

Frequency analysis revealed that gamma band activities consistently achieved the highest MI values across both datasets. However, fear emotion in SEED-IV exhibited notably lower MI values across all frequency bands compared to other emotions, with its maximum value (0.5776 at Fp2 in gamma) substantially lower than corresponding values for sad (0.7062) and happy (0.6954) emotions. The consistent prominence of frontal midline structures across both datasets aligns with their established role in emotion regulation\cite{10684098}, while the increased importance of posterior regions in SEED-IV suggests enhanced visual processing demands for four-emotion classification. This comprehensive analysis demonstrates that HADUA learns neurophysiologically plausible features that align with established emotion processing networks.

\subsection{Sensitivity Analysis for Hyperparameters}
To evaluate the robustness of the proposed class-aware reweighting strategy, we conduct sensitivity experiments on two key hyperparameters: the temperature coefficient $\tau$ and the alignment strength $\alpha$. These parameters jointly influence the sharpness of probability adjustment and the degree of class-distribution regularization.

\subsubsection{Sensitivity Analysis for Batch Size and Training Epoch}
\begin{figure}[t]
	\centering
	\includegraphics[width=\linewidth]{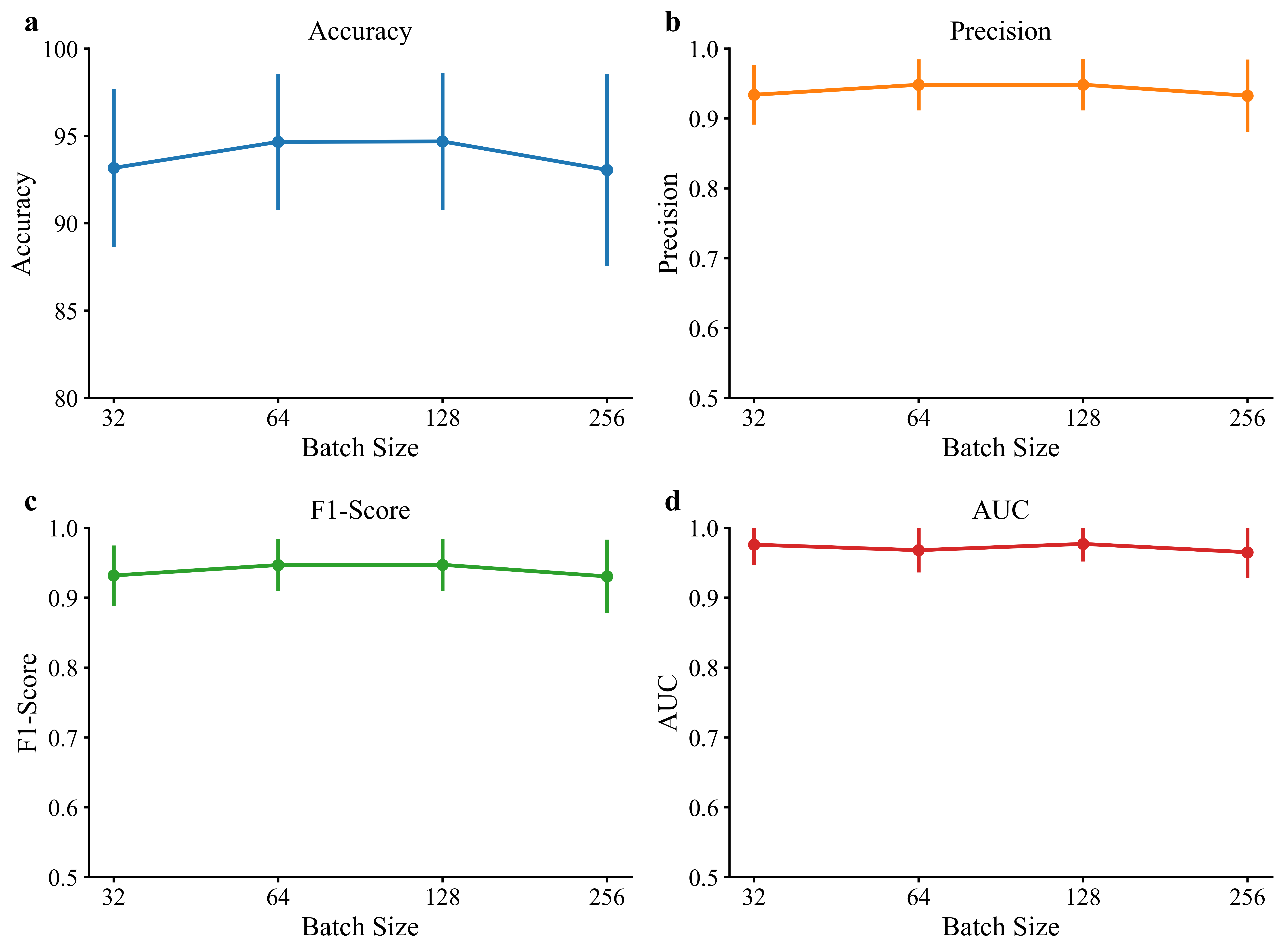}
	\caption{Sensitivity of HADUA to batch size.}
	\label{fig:heatmap_batch-size}
\end{figure}
\begin{figure}[t]
	\centering
	\includegraphics[width=\linewidth]{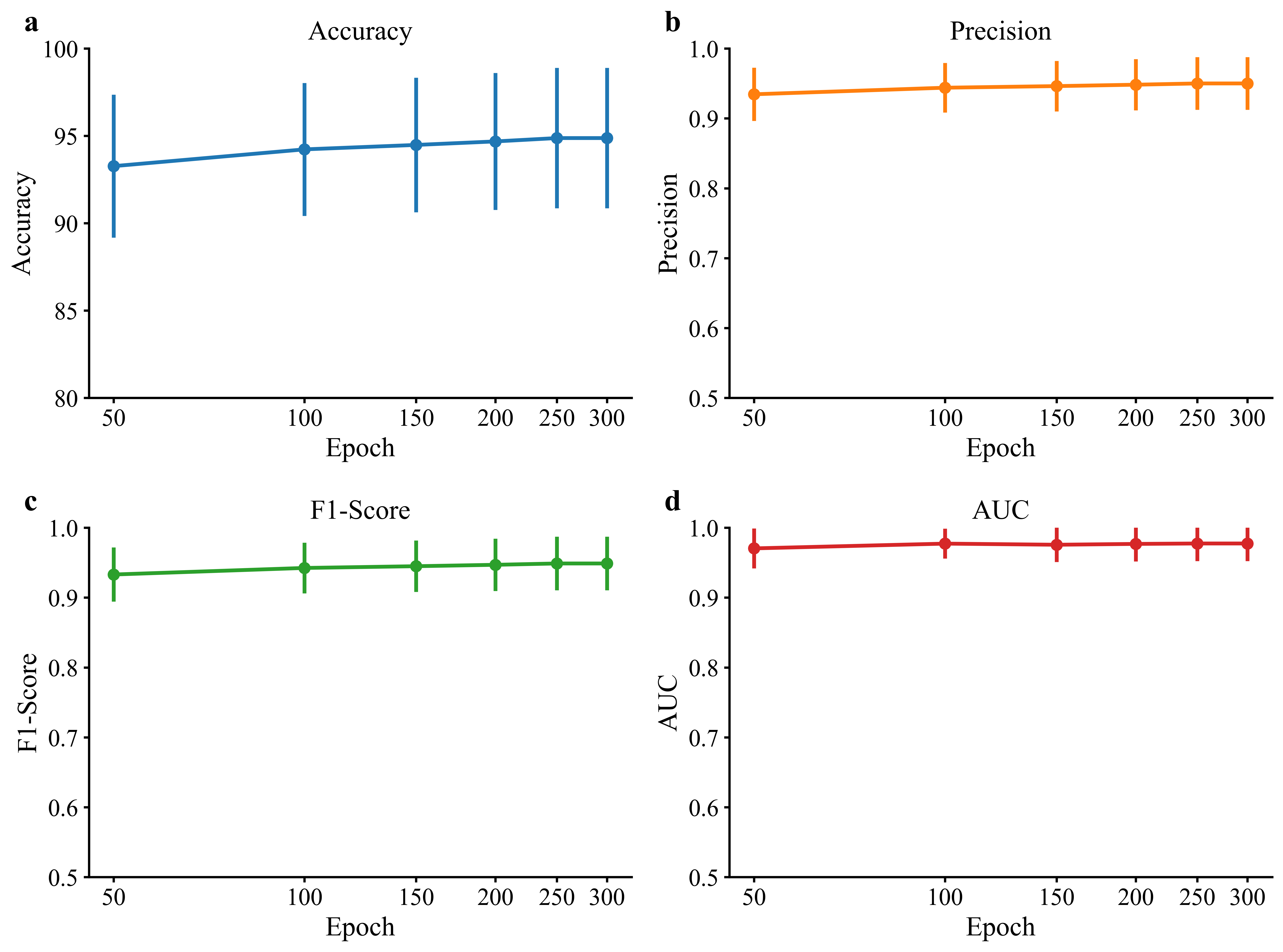}
	\caption{Sensitivity of HADUA to epoch.}
	\label{fig:heatmap_epoch}
\end{figure}
\begin{figure}[t]
	\centering
	\includegraphics[width=\linewidth]{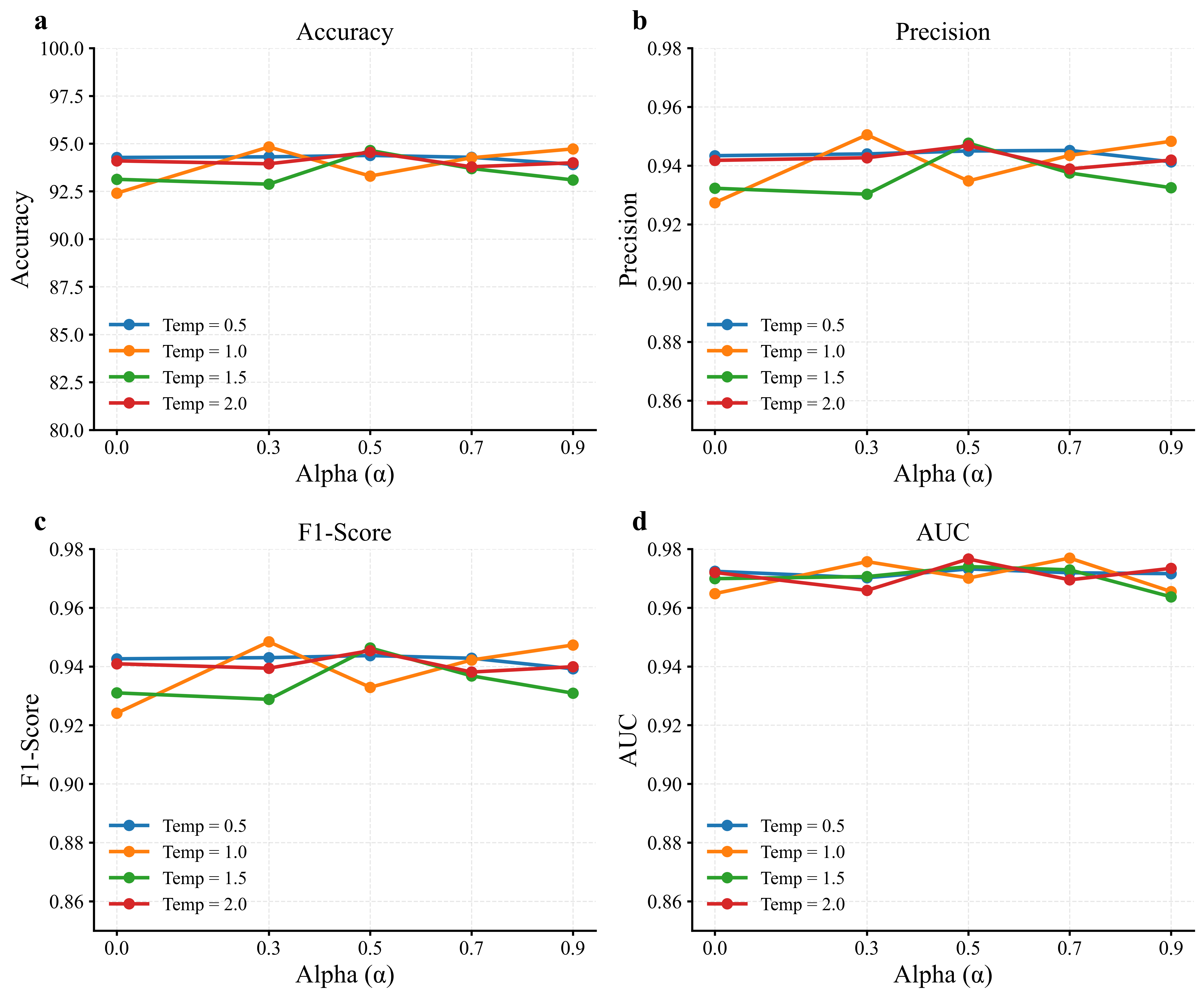}
	\caption{Sensitivity analysis of HADUA with respect to the temperature parameter $\tau$ and the alignment strength $\alpha$.}
	\label{fig:heatmap_accuracy}
\end{figure}
To better demonstrate the stability and effectiveness of the proposed HADUA framework, the average accuracies across all 12 test subjects are visualized with respect to batch size and training epochs in Fig.~\ref{fig:heatmap_batch-size}. We evaluate the model under a comprehensive grid of hyperparameter settings: batch sizes $\in \{16, 32, 64, 128, 256\}$ and epochs $\in \{50, 100, 150, 200, 250, 300\}$. As shown in Fig.~\ref{fig:heatmap_epoch}, the model achieves the highest average accuracy of 93.60\% at batch size = 256 and 250 epochs, significantly outperforming smaller batch configurations (e.g., 91.34\% at batch size 16, 50 epochs). Notably, larger batch sizes (256) require more epochs to reach peak performance, but once the epoch count exceeds 200, further training yields negligible gains (250 vs. 300 epochs show identical 93.60\% accuracy). This indicates that the model becomes insensitive to extended training beyond a certain threshold, demonstrating strong stability. In contrast, smaller batch sizes (e.g., 16) converge rapidly within the first 100 epochs but plateau early, reflecting limited generalization capacity. The smooth transition from light to dark blue regions in the heatmap further confirms that accuracy improves monotonically and stably with increasing batch size and moderate epoch growth. Additionally, rapid performance gains are observed within the first 50–100 epochs across most configurations (especially for batch size 256), suggesting that the proposed framework converges efficiently and does not require excessive training. These results collectively validate the robustness of HADUA to hyperparameter variations and highlight batch size = 256, epochs = 250 as the optimal and stable configuration for deployment.

\subsubsection{Sensitivity to Temperature $\tau$}
As formulated in Eq.~(\ref{eq:ua_adjust}), the temperature parameter $\tau$ controls the degree of non-linearity in the alignment adjustment, while simultaneously regulating the smoothness of the Gaussian confidence weighting applied to target-domain predictions. Across all tested values $\tau \in\{0.5, 1.0, 1.5, 2.0\}$, the method exhibits stable performance. Moderate temperatures such as $\tau = 0.5$ and $\tau = 1.0$ consistently produce the best results, achieving accuracies above $94.2\%$. For example, $\tau = 0.5, \alpha = 0.5$ yields $94.38\%$, while $\tau = 1.0, \alpha = 0.3$ achieves the highest performance of $94.83\%$. When $\tau$ becomes excessively large ($\tau \ge 1.5$), the adjustment becomes overly smooth, weakening class contrast and causing mild accuracy degradation. Nevertheless, the overall variation remains small ($<2\%$), demonstrating the robustness of the Gaussian weighting mechanism.

\subsubsection{Sensitivity to Alignment Strength $\alpha$}
As defined in Eq.~(\ref{eq:alpha_schedule}), the alignment strength $\alpha$ determines the extent to which the predicted class distribution in the target domain is encouraged toward a uniform prior during adaptation. The results show a clear trend: when $\alpha = 0.0$, alignment is disabled, leading to biased pseudo-label distributions and inferior performance. Increasing $\alpha$ into a moderate range ($0.3 \le \alpha \le 0.5$) substantially improves accuracy, precision, and F1-score. Notably, configurations such as $\tau = 0.5, \alpha = 0.5$ and $\tau = 1.5, \alpha = 0.5$ achieve accuracies of $94.38\%$ and $94.64\%$, respectively. However, overly aggressive alignment ($\alpha = 0.9$) forces the pseudo-label distribution too strongly toward uniformity, disrupting genuine class-specific tendencies and producing a consistent performance drop across all temperatures.

Both $\tau$ and $\alpha$ exhibit smooth, unimodal response curves, and the overall performance remains stable across a wide hyperparameter range. These results confirm that the proposed UA and Gaussian confidence weighting are robust, easy to tune, and effective under broad hyperparameter settings, ensuring strong generalization in cross-domain EEG emotion recognition.

\subsection{Why HADUA Works? Effectiveness Analysis}
The core challenge in cross-subject multimodal emotion recognition lies in the intertwined difficulties of modality heterogeneity and inter-subject distribution shift. These issues do not exist in isolation; rather, they form a vicious cycle: inconsistent multimodal representations lead to noisy target-domain predictions, which corrupt pseudo-labels and ultimately destabilize distribution alignment. This self-reinforcing degradation is a well-known bottleneck in domain adaptation\cite{pan2009survey,kouw2019review}. To break this cycle, HADUA is designed as an integrated framework that simultaneously addresses three coupled aspects: multimodal representation learning, pseudo-label reliability, and distribution alignment. It enhances representation consistency across modalities, stabilizes pseudo-label optimization at the sample level, and balances pseudo-label statistics at the class level. This coordinated approach enables marginal and conditional distribution alignment to function in a stable, complementary manner, which is the fundamental reason for its robust and stable performance.

At the representation level, HADUA employs hierarchical attention to model both intra-modality temporal dependencies and inter-modality semantic interactions between EEG and eye-movement signals. Compared with naive feature concatenation or rigid correlation-based fusion, modality-specific self-attention captures structured dependencies within each modality, while cross-attention facilitates guided information integration across modalities. This design effectively alleviates two common problems in cross-subject multimodal learning: (i) noise amplification caused by unstructured mixing of heterogeneous signals, and (ii) spurious correlations introduced by over-aligning modalities with distinct physiological origins. Both issues have been reported in prior multimodal affective computing studies \cite{8283814,9395500,10575932}. As a result, the fused representations become more discriminative and consistent, leading to smoother and less biased target-domain predictions, which directly improves the quality of pseudo-labels used in subsequent adaptation.

Based on these improved predictions, HADUA further addresses the quantity–quality trade-off in pseudo-label learning. Under domain shift, hard confidence thresholding is often unstable: high thresholds discard a large number of informative target samples, especially in early training stages, while low thresholds introduce noisy pseudo-labels that impair optimization. This limitation has been extensively discussed in semi-supervised and self-training frameworks \cite{sohn2020fixmatchsimplifyingsemisupervisedlearning,chen2023softmatchaddressingquantityqualitytradeoff}. To overcome this issue, HADUA adopts a confidence-aware Gaussian weighting strategy that assigns continuous weights to target samples according to prediction confidence. This mechanism allows moderately confident samples to participate in training with controlled influence, increasing effective supervision while preventing low-quality pseudo-labels from dominating the learning process. Consequently, pseudo-label-based self-training becomes more stable and less sensitive to heuristic threshold choices. Despite improved sample-level reliability, pseudo-labels in the target domain may still exhibit severe class imbalance, as the model tends to over-predict ’easy’ categories while under-representing ‘hard’ ones. Such imbalance is particularly detrimental to class-conditional alignment methods, which rely on reliable estimation of class-wise statistics \cite{10506974,10509712}. To address this issue, HADUA introduces UA to softly regularize the predicted target-domain class distribution toward a uniform prior. This strategy suppresses the dominance of majority classes and ensures that each emotion category contributes meaningfully to conditional distribution alignment. Importantly, UA acts as a soft regularization rather than hard re-labeling, preserving semantic information while improving the stability and fairness of class-conditional statistics used in CMMD \cite{10506974,10819285}.

With both sample-level confidence weighting and class-level distribution balancing in place, marginal and conditional distribution alignment can operate synergistically. Marginal alignment reduces global domain discrepancy by matching overall feature distributions, while conditional alignment preserves class-specific semantics by aligning class-conditional embeddings. Without reliable and balanced pseudo-labels, conditional alignment can easily become biased or unstable, for example collapsing toward dominant classes or amplifying early prediction errors. HADUA prevents this degeneration by ensuring that the pseudo-labels fed into the CMMD loss are both confidence-controlled and class-balanced, enabling effective semantic alignment across domains without sacrificing discriminative power.

The effectiveness of HADUA is further supported by experimental results. Confusion matrices show that misclassifications are mainly confined to affectively adjacent emotion categories rather than systematic bias toward dominant classes, indicating balanced class discrimination under cross-subject settings. In addition, the reduced standard deviation of per-class accuracies reflects improved class-wise consistency. Feature visualization results demonstrate progressive formation of compact emotion clusters together with improved source–target overlap within each class, confirming that the learned representations become both discriminative and domain-invariant during training. These observations are consistent with the expected behavior of well-posed joint marginal–conditional alignment.


\section{Conclusion}
In this work, we introduced HADUA, a unified framework that simultaneously addresses multimodal heterogeneity and domain shift for cross-subject emotion recognition. By integrating hierarchical multimodal fusion, confidence-aware pseudo-label refinement, and joint distribution alignment, our method achieves stable representation learning and effective adaptation to unseen subjects. Benchmark evaluations demonstrate that HADUA outperforms existing approaches, delivering superior accuracy, improved class balance, and stronger generalization across subjects.
Looking ahead, we aim to extend this framework to more challenging adaptation scenarios, such as open-set settings, potentially through proxy-based label refinement or uncertainty quantification. Further exploration of richer multimodal setups and deployment in real-world affective interaction systems will be the key to advancing practical applications.

\section{References }
\balance
\bibliographystyle{unsrt}

\begin{thebibliography}{10}

\bibitem{911197}
R.~Cowie, E.~Douglas-Cowie, N.~Tsapatsoulis, G.~Votsis, S.~Kollias, W.~Fellenz,
  and J.G. Taylor.
\newblock Emotion recognition in human-computer interaction.
\newblock {\em IEEE Signal Processing Magazine}, 18(1):32--80, 2001.

\bibitem{YOO2006345}
Seung~Hee Yoo, David Matsumoto, and Jeffrey~A. LeRoux.
\newblock The influence of emotion recognition and emotion regulation on
  intercultural adjustment.
\newblock {\em International Journal of Intercultural Relations},
  30(3):345--363, 2006.

\bibitem{RAHMAN2021104696}
Md.~Mustafizur Rahman, Ajay~Krishno Sarkar, Md.~Amzad Hossain, Md.~Selim
  Hossain, Md.~Rabiul Islam, Md.~Biplob Hossain, Julian~M.W. Quinn, and
  Mohammad~Ali Moni.
\newblock Recognition of human emotions using eeg signals: A review.
\newblock {\em Computers in Biology and Medicine}, 136:104696, 2021.

\bibitem{9395500}
Wei Liu, Jie-Lin Qiu, Wei-Long Zheng, and Bao-Liang Lu.
\newblock Comparing recognition performance and robustness of multimodal deep
  learning models for multimodal emotion recognition.
\newblock {\em IEEE Transactions on Cognitive and Developmental Systems},
  14(2):715--729, 2022.

\bibitem{6944757}
Wei-Long Zheng, Bo-Nan Dong, and Bao-Liang Lu.
\newblock Multimodal emotion recognition using eeg and eye tracking data.
\newblock In {\em 2014 36th Annual International Conference of the IEEE
  Engineering in Medicine and Biology Society}, pages 5040--5043, 2014.

\bibitem{8283814}
W.~Zheng, W.~Liu, Y.~Lu, B.~Lu, and A.~Cichocki.
\newblock Emotionmeter: A multimodal framework for recognizing human emotions.
\newblock {\em IEEE Transactions on Cybernetics}, pages 1--13, 2018.

\bibitem{10.3389/fnins.2018.00162}
Xiang Li, Dawei Song, Peng Zhang, Yazhou Zhang, Yuexian Hou, and Bin Hu.
\newblock Exploring eeg features in cross-subject emotion recognition.
\newblock {\em Frontiers in Neuroscience}, Volume 12 - 2018, 2018.

\bibitem{10.1145/3664647.3681579}
Wuliang Huang, Yiqiang Chen, Xinlong Jiang, Chenlong Gao, Qian Chen, Teng
  Zhang, Bingjie Yan, Yifan Wang, and Jianrong Yang.
\newblock Correlation-driven multi-modality graph decomposition for
  cross-subject emotion recognition.
\newblock In {\em Proceedings of the 32nd ACM International Conference on
  Multimedia}, MM '24, page 2272–2281, New York, NY, USA, 2024. Association
  for Computing Machinery.

\bibitem{10575932}
Chuangquan Chen, Zhencheng Li, Kit~Ian Kou, Jie Du, Chen Li, Hongtao Wang, and
  Chi-Man Vong.
\newblock Comprehensive multisource learning network for cross-subject
  multimodal emotion recognition.
\newblock {\em IEEE Transactions on Emerging Topics in Computational
  Intelligence}, 9(1):365--380, 2025.

\bibitem{pan2009survey}
Sinno~Jialin Pan and Qiang Yang.
\newblock A survey on transfer learning.
\newblock {\em IEEE Transactions on knowledge and data engineering},
  22(10):1345--1359, 2009.

\bibitem{kouw2019review}
Wouter~M Kouw and Marco Loog.
\newblock A review of domain adaptation without target labels.
\newblock {\em IEEE transactions on pattern analysis and machine intelligence},
  43(3):766--785, 2019.

\bibitem{10.5555/2832249.2832411}
Yifei Lu, Wei-Long Zheng, Binbin Li, and Bao-Liang Lu.
\newblock Combining eye movements and eeg to enhance emotion recognition.
\newblock In {\em Proceedings of the 24th International Conference on
  Artificial Intelligence}, IJCAI'15, page 1170–1176. AAAI Press, 2015.

\bibitem{wu2020transfer}
Dongrui Wu, Yifan Xu, and Bao-Liang Lu.
\newblock Transfer learning for eeg-based brain--computer interfaces: A review
  of progress made since 2016.
\newblock {\em IEEE Transactions on Cognitive and Developmental Systems},
  14(1):4--19, 2020.

\bibitem{8882370}
Jinpeng Li, Shuang Qiu, Changde Du, Yixin Wang, and Huiguang He.
\newblock Domain adaptation for eeg emotion recognition based on latent
  representation similarity.
\newblock {\em IEEE Transactions on Cognitive and Developmental Systems},
  12(2):344--353, 2020.

\bibitem{10506974}
Xinrong Gong, C.~L.~Philip Chen, Bin Hu, and Tong Zhang.
\newblock Ciabl: Completeness-induced adaptative broad learning for
  cross-subject emotion recognition with eeg and eye movement signals.
\newblock {\em IEEE Transactions on Affective Computing}, 15(4):1970--1984,
  2024.

\bibitem{10819285}
Magdiel Jiménez-Guarneros, Gibran Fuentes-Pineda, and Jonas Grande-Barreto.
\newblock Mmda: A multimodal and multisource domain adaptation method for
  cross-subject emotion recognition from a signals.
\newblock {\em IEEE Transactions on Computational Social Systems}, pages 1--14,
  2024.

\bibitem{HANGLOO2025130827}
Sakshini Hangloo and Bhavna Arora.
\newblock Multimodal fusion techniques: Review, data representation,
  information fusion, and application areas.
\newblock {\em Neurocomputing}, 649:130827, 2025.

\bibitem{electronics12153325}
Yundong Li, Longxia Guo, and Yizheng Ge.
\newblock Pseudo labels for unsupervised domain adaptation: A review.
\newblock {\em Electronics}, 12(15), 2023.

\bibitem{sohn2020fixmatchsimplifyingsemisupervisedlearning}
Kihyuk Sohn, David Berthelot, Chun-Liang Li, Zizhao Zhang, Nicholas Carlini,
  Ekin~D. Cubuk, Alex Kurakin, Han Zhang, and Colin Raffel.
\newblock Fixmatch: Simplifying semi-supervised learning with consistency and
  confidence, 2020.

\bibitem{chen2023softmatchaddressingquantityqualitytradeoff}
Hao Chen, Ran Tao, Yue Fan, Yidong Wang, Jindong Wang, Bernt Schiele, Xing Xie,
  Bhiksha Raj, and Marios Savvides.
\newblock Softmatch: Addressing the quantity-quality trade-off in
  semi-supervised learning, 2023.

\bibitem{10509712}
Yi~Yang, Ze~Wang, Wei Tao, Xucheng Liu, Ziyu Jia, Boyu Wang, and Feng Wan.
\newblock Spectral-spatial attention alignment for multi-source domain
  adaptation in eeg-based emotion recognition.
\newblock {\em IEEE Transactions on Affective Computing}, 15(4):2012--2024,
  2024.

\bibitem{10.3389/fncom.2016.00085}
Juan-Miguel López-Gil, Jordi Virgili-Gomá, Rosa Gil, Teresa Guilera, Iolanda
  Batalla, Jorge Soler-González, and Roberto García.
\newblock Method for improving eeg based emotion recognition by combining it
  with synchronized biometric and eye tracking technologies in a non-invasive
  and low cost way.
\newblock {\em Frontiers in Computational Neuroscience}, Volume 10 - 2016,
  2016.

\bibitem{10731546}
Wei-Bang Jiang, Xuan-Hao Liu, Wei-Long Zheng, and Bao-Liang Lu.
\newblock Seed-vii: A multimodal dataset of six basic emotions with continuous
  labels for emotion recognition.
\newblock {\em IEEE Transactions on Affective Computing}, pages 1--16, 2024.

\bibitem{Liu2016}
W.~Liu, W.-L. Zheng, and B.-L. Lu.
\newblock Emotion recognition using multimodal deep learning.
\newblock In {\em Proc. 23rd Int. Conf. Neural Inf. Process. (ICONIP)}, pages
  521--529, 2016.

\bibitem{10.3389/fnins.2023.1234162}
Baole Fu, Chunrui Gu, Ming Fu, Yuxiao Xia, and Yinhua Liu.
\newblock A novel feature fusion network for multimodal emotion recognition
  from eeg and eye movement signals.
\newblock {\em Frontiers in Neuroscience}, Volume 17 - 2023, 2023.

\bibitem{Qiu2018}
J.-L. Qiu, W.~Liu, and B.-L. Lu.
\newblock Multi-view emotion recognition using deep canonical correlation
  analysis.
\newblock In {\em Proc. Int. Conf. Neural Inf. Process.}, pages 221--231, Cham,
  Switzerland, 2018. Springer.

\bibitem{ZHOU2022108889}
Sijin Zhou, Dongmin Huang, Cheng Liu, and Dazhi Jiang.
\newblock Objectivity meets subjectivity: A subjective and objective feature
  fused neural network for emotion recognition.
\newblock {\em Applied Soft Computing}, 122:108889, 2022.

\bibitem{10643252}
Zequan Lian, Tao Xu, Zhen Yuan, Junhua Li, Nitish Thakor, and Hongtao Wang.
\newblock Driving fatigue detection based on hybrid electroencephalography and
  eye tracking.
\newblock {\em IEEE Journal of Biomedical and Health Informatics},
  28(11):6568--6580, 2024.

\bibitem{10.1145/3746027.3755459}
Ziyi Li, Wei-Long Zheng, and Bao-Liang Lu.
\newblock Multimodal emotion recognition with missing modality via a unified
  multi-task pre-training framework.
\newblock In {\em Proceedings of the 33rd ACM International Conference on
  Multimedia}, MM '25, page 5717–5725, New York, NY, USA, 2025. Association
  for Computing Machinery.

\bibitem{AN2025113613}
Yanling An, Shaohai Hu, Shuaiqi Liu, Xinrui Wang, Zhihui Gu, and Yudong Zhang.
\newblock Lgdaan-nets: A local and global domain adversarial attention neural
  networks for eeg emotion recognition.
\newblock {\em Knowledge-Based Systems}, 318:113613, 2025.

\bibitem{10938180}
Qi~Zhu, Ting Zhu, Lunke Fei, Chuhang Zheng, Wei Shao, David Zhang, and Daoqiang
  Zhang.
\newblock Multi-modal cross-subject emotion feature alignment and recognition
  with eeg and eye movements.
\newblock {\em IEEE Transactions on Affective Computing}, 16(3):2102--2115,
  2025.

\bibitem{qiu2024review}
Sen Qiu, Yongtao Chen, Yulin Yang, Pengfei Wang, Zhelong Wang, Hongyu Zhao,
  Yuntong Kang, and Ruicheng Nie.
\newblock A review on semi-supervised learning for eeg-based emotion
  recognition.
\newblock {\em Information Fusion}, 104:102190, 2024.

\bibitem{9904937}
Guangyi Zhang, Vandad Davoodnia, and Ali Etemad.
\newblock Parse: Pairwise alignment of representations in semi-supervised eeg
  learning for emotion recognition.
\newblock {\em IEEE Transactions on Affective Computing}, 13(4):2185--2200,
  2022.

\bibitem{10570465}
Baole Fu, Wenhao Chu, Chunrui Gu, and Yinhua Liu.
\newblock Cross-modal guiding neural network for multimodal emotion recognition
  from eeg and eye movement signals.
\newblock {\em IEEE Journal of Biomedical and Health Informatics},
  28(10):5865--5876, 2024.

\bibitem{song2018eeg}
Tengfei Song, Wenming Zheng, Peng Song, and Zhen Cui.
\newblock Eeg emotion recognition using dynamical graph convolutional neural
  networks.
\newblock {\em IEEE Transactions on Affective Computing}, 11(3):532--541, 2018.

\bibitem{ZHU2024124001}
Mu~Zhu, Qingzhou Wu, Zhongli Bai, Yu~Song, and Qiang Gao.
\newblock Eeg-eye movement based subject dependence, cross-subject, and
  cross-session emotion recognition with multidimensional homogeneous encoding
  space alignment.
\newblock {\em Expert Systems with Applications}, 251:124001, 2024.

\bibitem{article}
Magdiel Jiménez-Guarneros and Gibran Fuentes-Pineda.
\newblock Cfda-csf: A multi-modal domain adaptation method for cross-subject
  emotion recognition.
\newblock {\em IEEE Transactions on Affective Computing}, PP, 01 2024.

\bibitem{JIMENEZGUARNEROS2025113238}
Magdiel Jiménez-Guarneros and Gibran Fuentes-Pineda.
\newblock Multi-modal supervised domain adaptation with a multi-level alignment
  strategy and consistent decision boundaries for cross-subject emotion
  recognition from eeg and eye movement signals.
\newblock {\em Knowledge-Based Systems}, 315:113238, 2025.

\bibitem{zheng2015investigating}
Wei-Long Zheng and Bao-Liang Lu.
\newblock Investigating critical frequency bands and channels for {EEG}-based
  emotion recognition with deep neural networks.
\newblock {\em IEEE Transactions on Autonomous Mental Development},
  7(3):162--175, 2015.

\bibitem{koelstra2011deap}
Sander Koelstra, Christian Muhl, Mohammad Soleymani, Jong-Seok Lee, Ashkan
  Yazdani, Touradj Ebrahimi, Thierry Pun, Anton Nijholt, and Ioannis Patras.
\newblock Deap: A database for emotion analysis; using physiological signals.
\newblock {\em IEEE transactions on affective computing}, 3(1):18--31, 2011.

\bibitem{chen2024comprehensive}
Chuangquan Chen, Zhencheng Li, Kit~Ian Kou, Jie Du, Chen Li, Hongtao Wang, and
  Chi-Man Vong.
\newblock Comprehensive multisource learning network for cross-subject
  multimodal emotion recognition.
\newblock {\em IEEE Transactions on Emerging Topics in Computational
  Intelligence}, 9(1):365--380, 2024.

\bibitem{10684098}
Hua Yang, C.~L.~Philip Chen, Bianna Chen, and Tong Zhang.
\newblock Improving the interpretability through maximizing mutual information
  for eeg emotion recognition.
\newblock {\em IEEE Transactions on Affective Computing}, 16(2):744--757, 2025.

\end{thebibliography}

\end{document}